Working document – HAL-ABXO4 – May 2019

# The variable properties of solid solutions $A_{1-x}B_xXO_4$: Tunable photoluminescence in the case of $Sr_{1-x}Pb_xWO_4$ series.

**J-R. Gavarri (1, *)**

**F. Guinneton (1), M. Arab (1), J-C. Valmalette (1), S. Villain (1), A. Hallaoui (2), A. Taoufyq (2), B.Bakiz (2), and A. Benlhachemi (2)**

(1) *University of Toulon, Aix-Marseille Univ., CNRS 7334, IM2NP, BP 20132, 83957 La Garde Cedex, France.*
(2) *Laboratory Materials & Environment (LME), Faculty of Sciences, University Ibn Zohr, Agadir, Morocco.*

(*) *To whom correspondence must be addressed: e-mail=gavarri.jr@univ-tln.fr & gavarri.jr@gmail.com*

## Abstract.

Over the ten past years, various experimental studies of solid solutions $A_{1-x}B_xXO_4$ (e.g. A = Sr, B = Pb, X = W) with scheelite structures have evidenced correlations between structural, vibrational modifications due to chemical substitution, and increasing photoluminescence intensities under UV or X-ray excitation. We propose a simple semi-empirical approach based on local zones with different compositions, allowing simulating the variations of structural, vibrational and photoluminescence characteristics, in the full composition range $0 \leq x \leq 1$. The structural characteristics are cell parameters, cell distortions or crystallite size effect, Debye-Waller factors, Raman shifts characterizing vibrations and photoluminescence signals under UV or X-ray excitations. Each property is assumed to be represented by a non-linear function Y(x) depending on composition x and on local microstructural disorder. To illustrate this approach based on the coexistence of local zones with different compositions, we have fitted the Y(x) function to experimental data, which allowed us determining the significant parameters characteristic of the series with A=Sr, B=Pb and X = W. These parameters deliver a new microstructural interpretation of the increasing photoluminescence intensities observed for intermediate composition x in solid solutions. A generalization of this approach to other series of solid solutions is quite possible.

## 1. Introduction.

**General objectives.** In the general framework of materials studies that can be used in technologies for radiation detection (sensors) or low-cost lighting (LED's), diversified types of photoluminescent materials susceptible to be used in radiation sensors or low-cost lighting (LED's) were systematically studied by the past: classically, in the literature many studies focused on pure compounds (pure tungstate, molybdates …) and doped compounds (rare earth doped materials). More recently, solid solution studies have been of great interest to the scientific community concerned with luminescence for industrial purposes.

**Scheelite structures.** The tungstates and molybdates $AXO_4$ with A = Ca, Sr, Ba, Pb, Cd… are well known for their potential applications: luminescence for detection, low cost lighting, photocatalysis [1-12]. The synthesis conditions or thermal treatments played a prominent role in the observed final properties. Recently, various series of solid solutions of photoluminescent materials were synthesized and studied in our laboratories: $Ca_{(1-x)}Cd_xWO_4$ [13], $Sr_{(1-x)}Pb_xWO_4$ [14], $Sr_{(1-x)}Pb_xMoO_4$ [15], $Ba_{(1-x)}Pb_xWO_4$ [16]. In each of these series of solid solutions, the structural, electrical, vibrational and photoluminescence properties were studied as a function of composition x. The effect of disorder was observed through perturbations of these properties. A systematic increase in photoluminescent emissions has been observed for certain composition x.

**Abnormal photoluminescence. Figure 1** reports the variations of these PL emissions in two examples previously published [14, 15, 17]. In the case of $Sr_{(1-x)}Pb_xWO_4$, $Sr_{(1-x)}Pb_xMoO_4$ solid solutions, maxima of intensities were observed for x values close to $x_{max}$ =0.3 under X-Ray excitation, and close to $x_{max}$ = 0.7 under UV excitation. For these two tungstates and molybdates series, the PL emissions under UV excitation were analyzed in Hallaoui's thesis [17], and their characteristics are shortly reported here. Under UV excitation, the maximum of intensity occurred close to $x_{max}$ = 0.7 for $Ca_{(1-x)}Cd_xWO_4$ (see [13]). In the case of barium lead tungstate [16], the maximum of PL intensity was located close to $x_{max}$ = 0.5. All these materials present scheelite structures with $WO_4$ tetrahedral groups, except the $Ca_{(1-x)}Cd_xWO_4$ series that crystallizes in scheelite structure for x < 0.6, and in wolframite structure (with $WO_6$ octahedral groups) for x> 0.7.

However, up to date, the correlations between the modified properties of solid solutions and the substitution were never clearly established: the departure from ideal linear behaviors was never described as a function of composition, and/or of synthesis conditions.

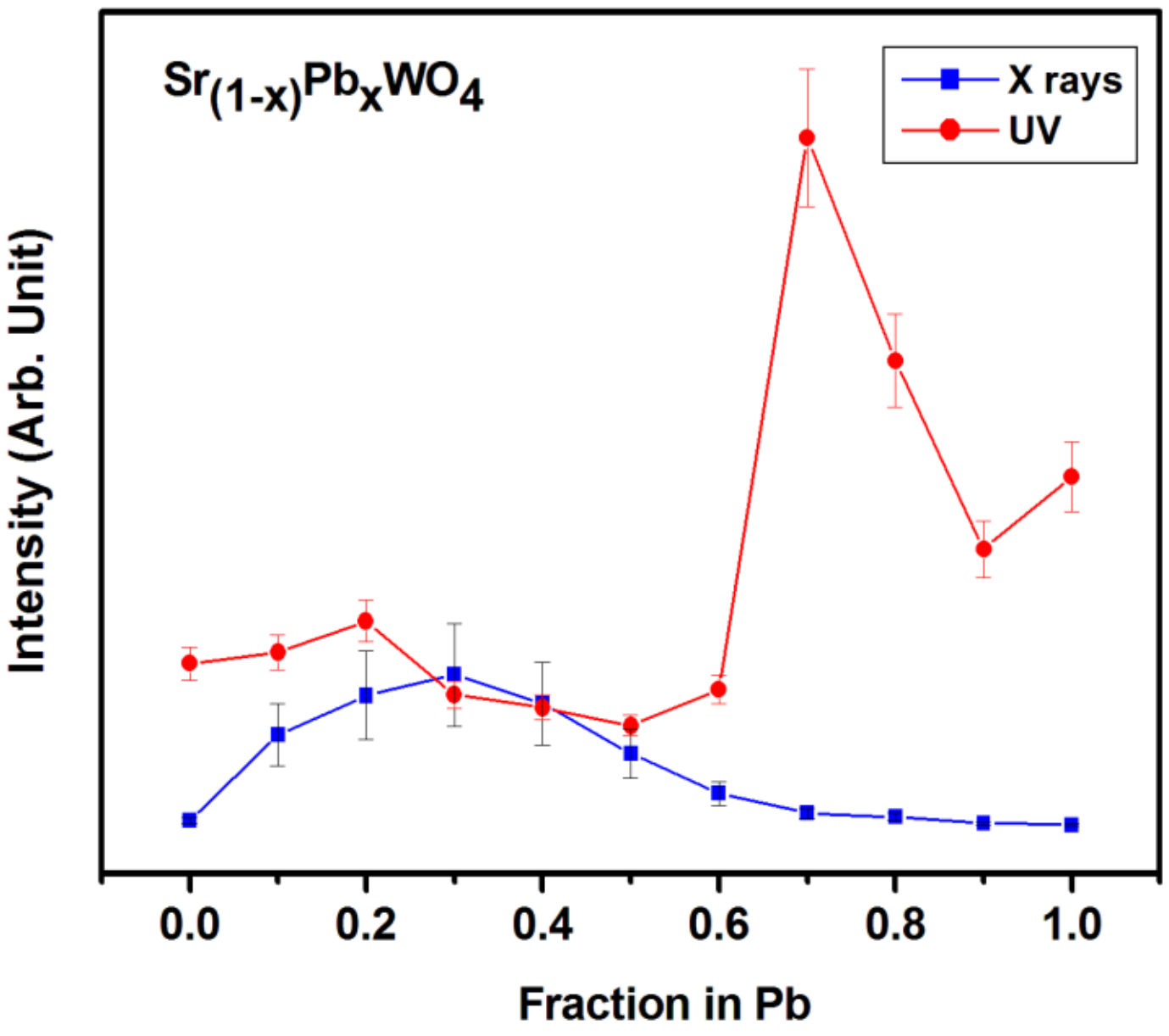

**Fig. 1a: Variations of PL emission intensities under X-ray or UV excitations, in the cases of $A_{1-x}B_xXO_4$ solid solutions with (A, B, X) = (Sr, Pb, W). Results from ref. [14, 15, 17].**

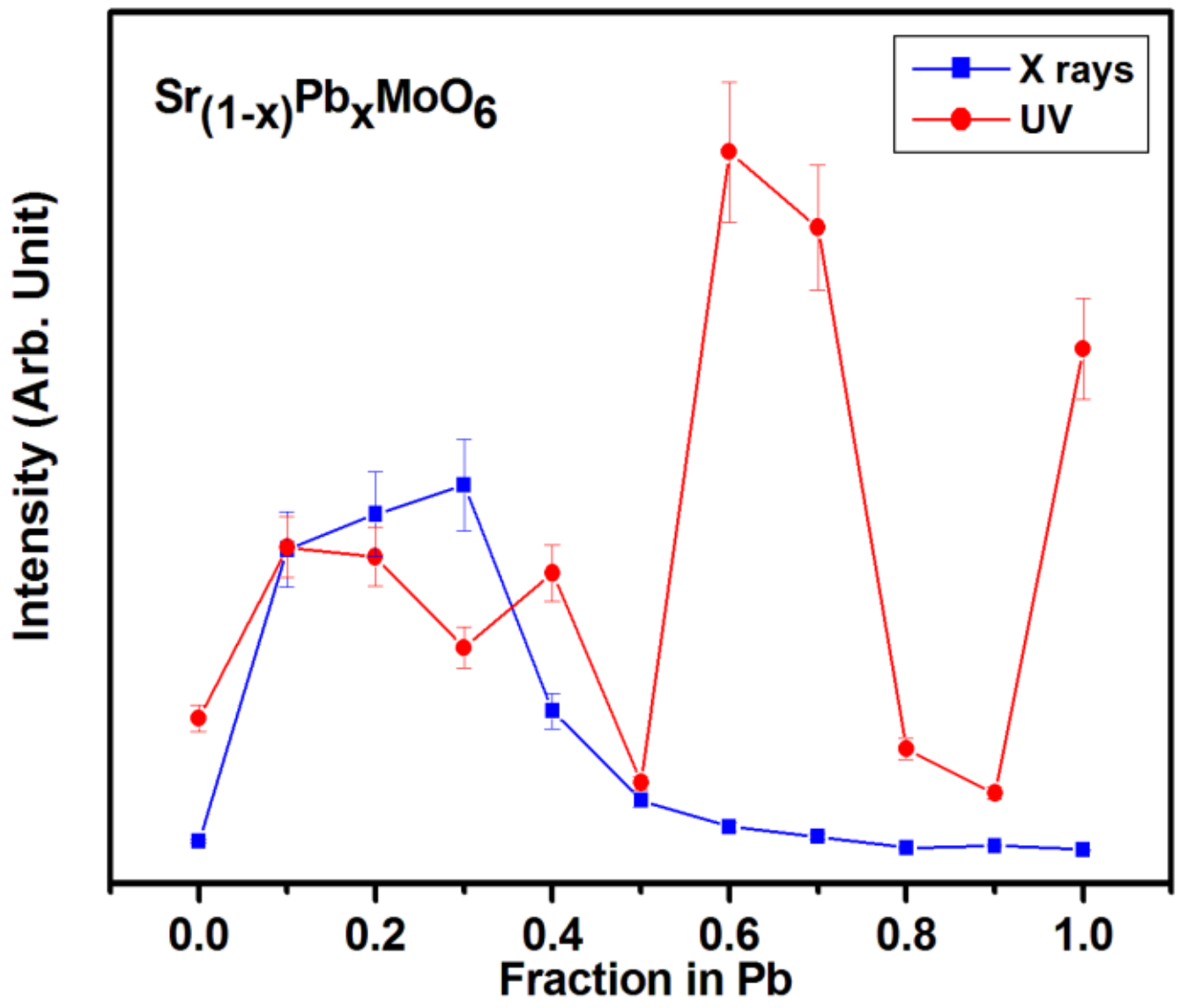

**Fig. 1b: Variations of PL emission intensities under X-ray or UV excitations, in the cases of $A_{1-x}B_xXO_4$ solid solutions with (A, B, X) = (Sr, Pb, Mo). Results from ref. [14, 15, 17].**

In this study, we present and test a semi-empirical model allowing simulating the variations of structural, vibrational, photoluminescence properties of substituted materials expressed in the form $\mathbf{A_{1-x}B_xXO_4}$. The general objective could be to propose specific parameters describing the observed evolutions of physical properties in the case of disordered solid solutions.

## 2. Methodology: disorder in solid solutions.

### 2.1. Disorder and local structures.

The solid solutions are assumed to be constituted of disordered distributions of bivalent cations $A^{2+}$ and $B^{2+}$ immersed in a distribution of $XO_4$ "molecular" groups. In solid solutions, two categories of defects can be juxtaposed:

(i) bivalent cations statistically distributed in the lattice, inducing local distortions due to different chemical bonds A-O and B-O and different interactions A-O-X-O-B: the substitution induces also modifications in energy levels and in band gaps;

(ii) intrinsic defects (e.g., cation and anion vacancies) inducing variable distortions in the lattice, resulting from the synthesis conditions and thermal treatments.

These different natures of disorder can play a drastic role in modifications of physical and chemical properties. It should also be noted a change in the morphologies and sizes of crystallites as the composition x varies, resulting from the different chemical bonds A-O and B-O. These changes may also affect the macroscopic properties of the material.

We propose to describe the composition dependence of structural or physical characteristics, by assuming that the properties of disordered solid solutions $A_{1-x}B_xXO_4$ can be conditioned not only by the distribution of A-O and B-O chemical bonds in the lattice, but also by the local distortions of $AO_8$ and $BO_8$ polyhedra associated with the formation of vacancies. These "perturbations" would involve distortions of $XO_4$ tetrahedra and formation of specific defects $[XO_3V_O]$, $V_O$ being an oxygen vacancy. Our central hypothesis would be that the crystal lattice could be constituted of juxtaposition of "AA=$AXO_4$", "BB=$BXO_4$" zones with intermediate mix zones "AB= $A_{1-y}B_yXO_4$"where the composition y differs from x, in which the $XO_4$ groups should be linked to A and B cations, giving rise to A-O-X-O-B interactions.

### 2.2. Simulation functions $\mathbf{Y_{ABXO4}(x)}$.

The main objective of the semi-empirical approach resides in fitting a simulation function noted $\mathbf{Y_{ABXO4}(x)}$ to experimental values of a property noted $\mathbf{P_{ABXO4}(x)}$ resulting from a given synthesis process including thermal treatments. This property depends on both defects and substitution.

**Microstructural hypotheses.** To represent such a macroscopic property of a material, we can assume that an individual crystallite is necessarily constituted of a juxtaposition of small-sized local zones, comparable to "local" nanophases $AXO_4$ (AA), $BXO_4$ (BB) and $A_{1-y}B_yXO_4$ (AB). Each crystallite would be represented by the formula $A_{1-x}B_xXO_4$, while the local zones AB would be characterized by a local composition y<x.

More precisely, the distribution of local zones (AA=$AXO_4$, BB=$BXO_4$, AB= $A_{1-y}B_y XO_4$) can be conditioned by the basic relationship relative to the nominal composition:

$$[\mathbf{A_{1-x}B_xXO_4}] = \alpha.\ [\mathbf{AXO_4}] + \beta.\ [\mathbf{BXO_4}] + \gamma.\ [\mathbf{A_{1-y}B_y XO_4}] \qquad \mathbf{(1)}$$

with $\alpha + \beta + \gamma = 1$ and $\mathbf{y = (x - \beta)/\gamma}$ .

The $\alpha$ and $\beta$ coefficients (or $\beta$ and $\gamma$) can characterize the effective state of disorder of the material: they depend on composition and synthesis conditions (thermal treatments). In the following we try to express these coefficients as functions of composition x.

**How to define order or disorder in a simple way?** Let us recall that a total disorder of $AXO_4$ and $BXO_4$ units would require disordered sequences (in 3 space directions):

(**AXO4//BXO4BXO4// AXO4AXO4AXO4// BXO4BXO4 // AXO4AXO4**…..)

e.g. in materials with composition x= ½ ($A_{½}B_{½}XO_4$)., If the local zones diffract coherently, the typical cell parameter would be : $<d_{(Disorder)}> = ½ [d_{AXO4} + d_{BXO4}]$.
In other terms, this should correspond to the virtual existence of an average atom $M = A_{1-x}B_x$.
Conversely, a total long-range order would be associated with a periodic repetition of structural units (in 3 space directions) of the type (e.g. in the case of a typical solid solution $A_{1-x}B_xXO_4$ with x = ½):

**(AXO4//BXO4//AXO4//BXO4/AXO4//BXO4// ……)**

with a unique repetition distance $d_{(Order)} = 2<d_{(Disorder)}>$ (giving rise to a superstructure cell parameter). In other terms, only the AB local zones would occupy the totality of lattice.

In the case of total disorder, these local zones are necessarily limited in extension to allow diffraction from a unique average lattice. Each individual crystallite would be characterized by different A, B atoms occupying the same crystallographic average position, but being displaced from their average position. Due to their different sizes, these atoms would induce different bond lengths A-O, B-O and would provoke small displacements of X and O atoms from their average positions.

In addition, it should be recalled that, in certain systems, depending on synthesis and thermal treatments, a partial order can be defined with an order parameter characterizing the statistical occupation factors of A and B on their official sites. This concept could be associated with a superposition of the AA, BB and AB zones defined just above. Depending on thermal treatments, for a given composition x, these zones could vary in extension.

For low or high x values (e.g. $x<0.2$ or $x>0.8$), the AA or BB zones occupy the major part of the lattice, respectively: their major defects (mainly vacancies and distortions) are intrinsic defects depending on synthesis conditions (thermal treatments).
For intermediate values of x (e.g. close to x=0.5), the AB zones are the more extended or frequent ones: structural defects result from the A-O-X-O-B interactions (with distortions due to different cation sizes) and from synthesis conditions.
As the extension and number of zones vary with increasing x values, we assume that different distributions of point defects and distortions should be formed in the three zones (AA, BB, AB).

To represent a given modified property linked to one AA, BB or AB zone (noted as $Y_{(AA, BB, AB)}$), we can assume that, due to interactions with neighboring zones, each local property $Y_{(AA,}$ $_{BB,\ AB)}$will be degraded (or modified) through a statistical function G depending on x and on thermal treatments.

For each zone, we can assume that this function G can take the form of a gaussian function:

**AA Zone ($AXO_4$):** $\mathbf{G_{AA} = exp\ (-k_{AA}(x\text{-}x_{AA})^2\ )}$ **(2a)**

**BB Zone ($BXO_4$):** $\mathbf{G_{BB} = exp\ (-k_{BB}(x\text{-}x_{BB})^2\ )}$ **(2b)**

**AB Mix zone ($A_{1-y}B_yXO_4$):** $\mathbf{G_{AB} = exp\ (-k_{AB}(x\text{-}x_{AB})^2)}$ **(2c)**

Three typical configurations can be illustrated to represent a disordered solid solution: (a) A-rich phase with x close to 0; (b) A and B in equal composition, with x= 1/2; (c) B-rich phase with x close to 1.

(a) : AAAAA**B**AAAA**BBBB**AAAAAA**B**AAAAAA**B** (21A+7B)

(b) : AAA**BBB**AAAA**B**A**BBA**ABBB**ABA**AABBBB (14A+14B)

(c) : BBBBB**AA**BBBBB**AB**ABB**AB**AABBBBBBB (21B+7A)

The $x_{(AA,\ BB,\ AB)}$ compositions correspond with optimal values of G as x varies. The typical values would be: $x_{AA}$ = 1 (for x=0 $G_{AA}$=1), $x_{BB}$ = 1 (for x=1 $G_{BB}$=1) and $x_{AB}$ a value depending on the history of material, for which $G_{AB}$ = 1. For intermediate x values e.g. close to ½, $G_{AB}$ =1 with $x_{AB}$= ½.

The $k_{(AA,BB,AB)}$ parameters characterize the forms of the three G functions: in other terms, if we use the expressions $k_{(AA,BB,AB)} = 1/(\delta x_{(AA,\ BB,\ AB)})^2$, the δx's characterize the widths of the gaussian.

These functions can be justified by the existence of defects and distortions resulting from the synthesis process and including morphological modifications, vacancies …. and so on. During the formation of these zones with increasing x values, we can consider that the number of defects depends on the mutual interactions of AA, BB and AB zones. For small x values (e.g. x<0.2, AA zones being the more extended ones), the AB and BB zones could be strongly perturbed but in small number. However, the AA zones occupying the major part of lattice could be weakly perturbed. For high x values (e.g. x>0.8) the AA and AB zones become the minority with many defects, while the BB zones become the majority and could be weakly perturbed. For intermediate x values (e.g. 0.4<x<0.6), the AB zones are the majority and the defects are mainly due to disorder and distortions induced by substitution.

The final choice of functions used in our simulations will be with $x_{AA}$=0, $x_{BB}$=1 and $x_{AB}$ a parameter to be adapted:

$\mathbf{G_{AA} = exp\ (-k_{AA}(x)^2)}$ **(3a) ;** $\mathbf{G_{BB} = exp\ (-k_{BB}.(1\text{-}x)^2)}$ **(3b) ;** $\mathbf{G_{AB} = exp\ (-k_{AB}(x\text{-}x_{AB})^2)}$ **(3c)**

The effective seven parameters {$Y_{(AA,BB,AB)}$, $x_{(AB)}$ and $k_{(AA,BB,AB)}$} will have to be determined to fit calculated functions Y(calculated) to experimental data $P_{(observed)}$. At this step, we must be careful in interpreting the values of these adaptation parameters: given the limited precision of certain $P_{(observed)}$ data, they only give qualitative indications on the defects in each zone AA, BB and AB. In addition, these parameters may vary from one property to another, for the same set of materials.

The modifications of certain properties (cell parameters, lattice distortions, B factors, wavelengths of Raman bands, PL emission intensities …) will depend on these distributions of defects and distortions characterized by gaussian functions G (see above). In other terms, each contribution could be roughly proportional to the associated fraction (1-x) for AA zones, x for BB zones and x(1-x) for AB zones.

The resulting general formula characteristic of functions $Y_P(x)$ used to fit experimental data $P_{ABXO4}(x)$ should be:

$$\mathbf{Y_P(x) = Y_{AA}.\ (1\text{-}x).G_{AA}\ (x) + Y_{BB}.\ x.\ G_{BB}\ (x) + Y_{AB}\ x.(1\text{-}x).\ G_{AB}(x)} \qquad \mathbf{(4)}$$

This expression (3) can be applied to various properties but under specific conditions.

**Structural data acquisition**. In the case of polycrystalline solid solutions (e.g. $A_{1-x}B_xXO_4$), the acquisition of structural data requires the use of **Rietveld procedure**, applied to a series of compounds having the same space group and cell parameters slowly varying with composition x. To characterize a continuous variation in structural parameters, the extreme phases ($AXO_4$ and $BXO_4$) of a solid solution must be considered as pure “standard” phases used to determine the continuous variations of parameters. This requires a perfect structural knowledge of these standards. The diffraction profile analyses (from FWHM’s determinations) can give additional data on coherence lengths (L(x) and/or distortions effects ε(x)). The intensities of Bragg peaks can also deliver data on atom coordinates, vibration amplitudes, static displacements and occupation factors, despite the high difficulty to obtain accurate data on these last characteristics.

In the case of Raman spectroscopy, the vibration bands also give informations on average vibration modes, perturbation of profiles due to sizes or distortions effects.

In the case of photoluminescence intensities $I_{PL}(x)$ under X-ray or UV excitations, each zone AA, BB and AB is susceptible to have different populations of emitting centers, depending on both composition x and synthesis conditions (more precisely thermal treatments).

## 3. Results of simulation: application to the $Sr_{1-x}Pb_xWO_4$ series.

The semi-empirical model has been applied to the solid solution **$Sr_{1-x}Pb_xWO_4$**[14] synthesized under polycrystalline form by solid-state reaction, in the composition range $0 \leq x \leq 1$. Eleven samples were prepared using polycrystalline precursors $WO_3$, $SrCO_3$ and PbO. The optimized elaboration conditions were as follows (**Figure 2**): the reagents in stoichiometric proportions were thoroughly mixed and ground in an agate mortar for 15 min, then thermally treated at 600°C for 3 hours, in pure alumina crucible under air. Each sample was ground again thoroughly for 2 hours and then retreated at 1100°C for 4 hours, under air. The choice of elaboration conditions is crucial to allow comparative study of properties.

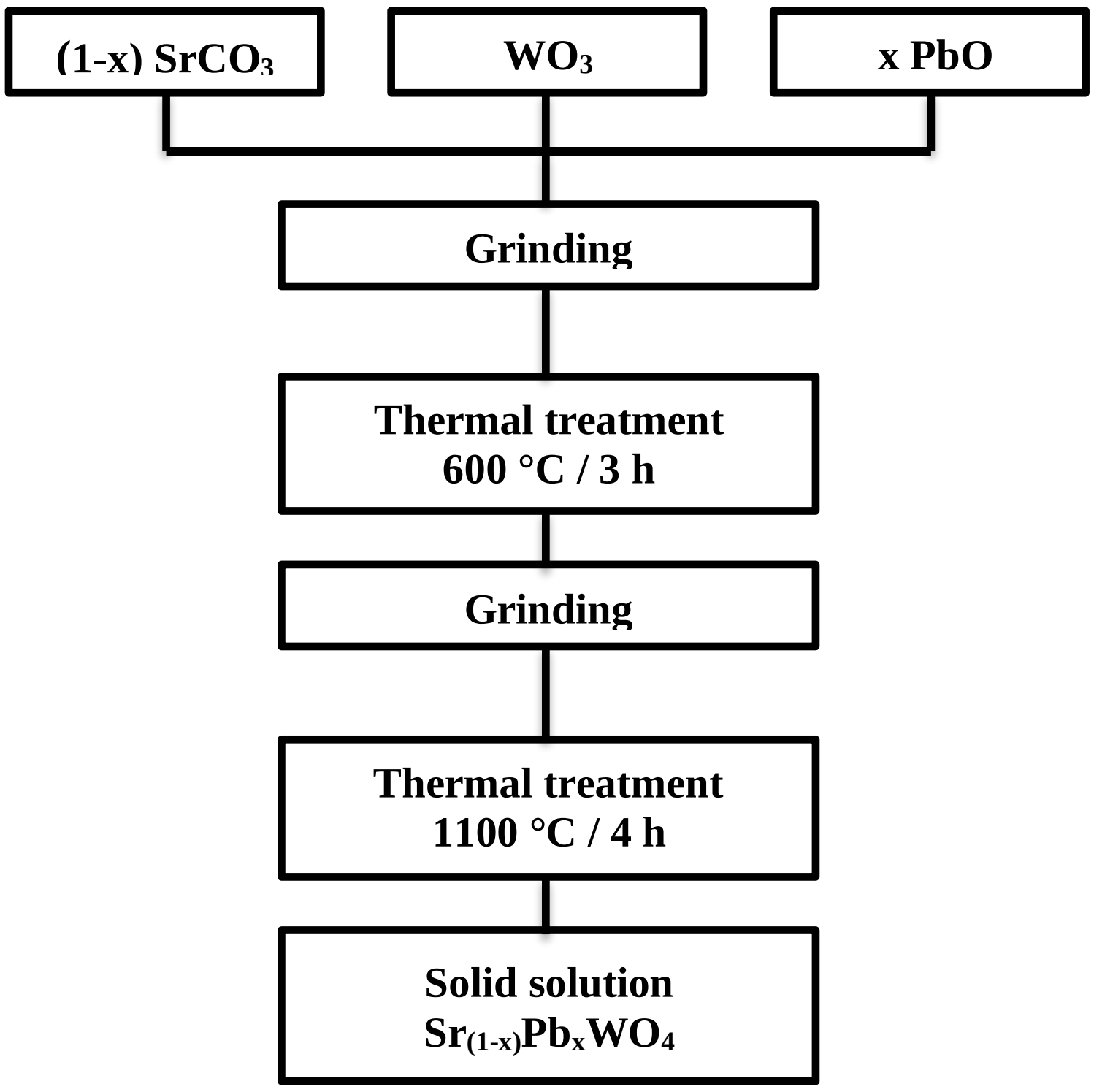

**Fig. 2: Synthesis conditions of polycrystalline samples in the case of $Sr_{1-x}Pb_xWO_4$ solid solution: unique process, same thermal treatment.**

Using Rietveld method, the structural data of all polycrystalline samples were refined [14] and crystal cell parameters exhibited a linear behavior as a function of x.

Correlations between cell parameters and Raman wavenumber shifts have clearly shown that substitution of Sr atoms by Pb atoms could modify W-O chemical bonds, through interactions Sr-O-W-O-Pb involving modifications in $WO_4$ groups due to covalence of Pb-O bonds. The structural modifications induced by the substitution are at the origin of the observed changes in the components of the photoluminescence (PL) emission spectra under X-ray or UV excitation. In the case of luminescence experiments under X-ray excitation, using the copper X-ray source of a standard diffractometer, four PL components were observed. They were typical characteristics of the whole solid solution: only two of them were sensitive to substitution rate, the other two seemed to be invariant in energy and characteristic of intrinsic defects in scheelite structure. The intensities of these bands exhibited a maximum for compositions close to x = 0.3 in the case of X-ray excitation or x=0.7 in the case of UV excitation. These two different compositions should result from various hypothetical physical factors:

- Emission centers increasing in numbers with the observed increasing number of defects due to substitution,

- As x increases, increasing Pb6s contributions in the [Pb6s - O2pσ*] hybrid orbitals located at the top of the valence band constituted of the O2p-W5d and O2pπ orbitals [3],
- Large or weak penetrations of X-ray or UV radiations, respectively, in the materials, with an increasing absorption of radiations due to increasing Pb content, involving two different emissions from bulk or surface of materials,
- Morphology modifications observed in the $Sr_{1-x}Pb_xWO_4$ series playing an additional role in the emissions under X-ray or UV excitations.

These evolutions in luminescence signals show that chemical substitution could play an interesting role to master and optimize the photoluminescence under X-ray or UV excitation, at least in polycrystalline materials.

In the following of the text, we note $P_{ABXO4}(x)$ a given property for a given composition x. This property is assumed to result from the coupling of substitution and synthesis conditions: it will not be strictly linearly correlated to each individual property of $AXO_4$ and $BXO_4$ phases. The simulated properties will be noted $Y_{ABXO4}(x)$.

### 3.1. Structural perturbations.

**Figure 3** reports a zoom of the X-ray diffraction patterns relative to polycrystalline samples of $Sr_{1-x}Pb_xWO_4$ with x ranging between 0 and 1. All diffraction patterns were exploited using Rietveld method (Fullprof software): see [14]. Fig. **3a** shows weak broadenings of Bragg peak profiles for intermediate compositions x. A detailed analysis of Bragg peak profiles shows that a progressive broadening of Bragg peaks is observed in the composition range $0.1<x<0.6$, with a maximum of deformation for the composition x=0.35. This perturbation of FWHM's is generally induced by size and distortion effects. However, the scanning electron microscopy analyses [14] showed that large crystallites, variable in sizes as composition x varied,were formed. This could suggest that a maximum of structural defects occurred close to the composition x=0.35.

#### 3.1.1. Cell parameters.

In **Fig. 3b,** we observe a linear increase of cell parameters a and c (noted as (a,c)) as a function of x. This is a clear indication that the solid solution is quasi ideal. The application of the model is thus trivial. We can consider that the Y(x) function for the cell parameters (a,c) can be expressed as follows:

$$\mathbf{Y(x) = (a,c) = (1-x).(a,c)_{AXO4} + x\,(a,c)_{BXO4} + x(1-x)\,(\alpha,\chi)_{AB}} \qquad \mathbf{(5)}$$

with $\mathbf{(\alpha,\chi)_{AB}=0}$. The couple of values $(a,c)_{AA}$ and $(a,c)_{BB}$ are in $10^{-10}$ m: (5.414/11.93) and (5.461/12.055) respectively.

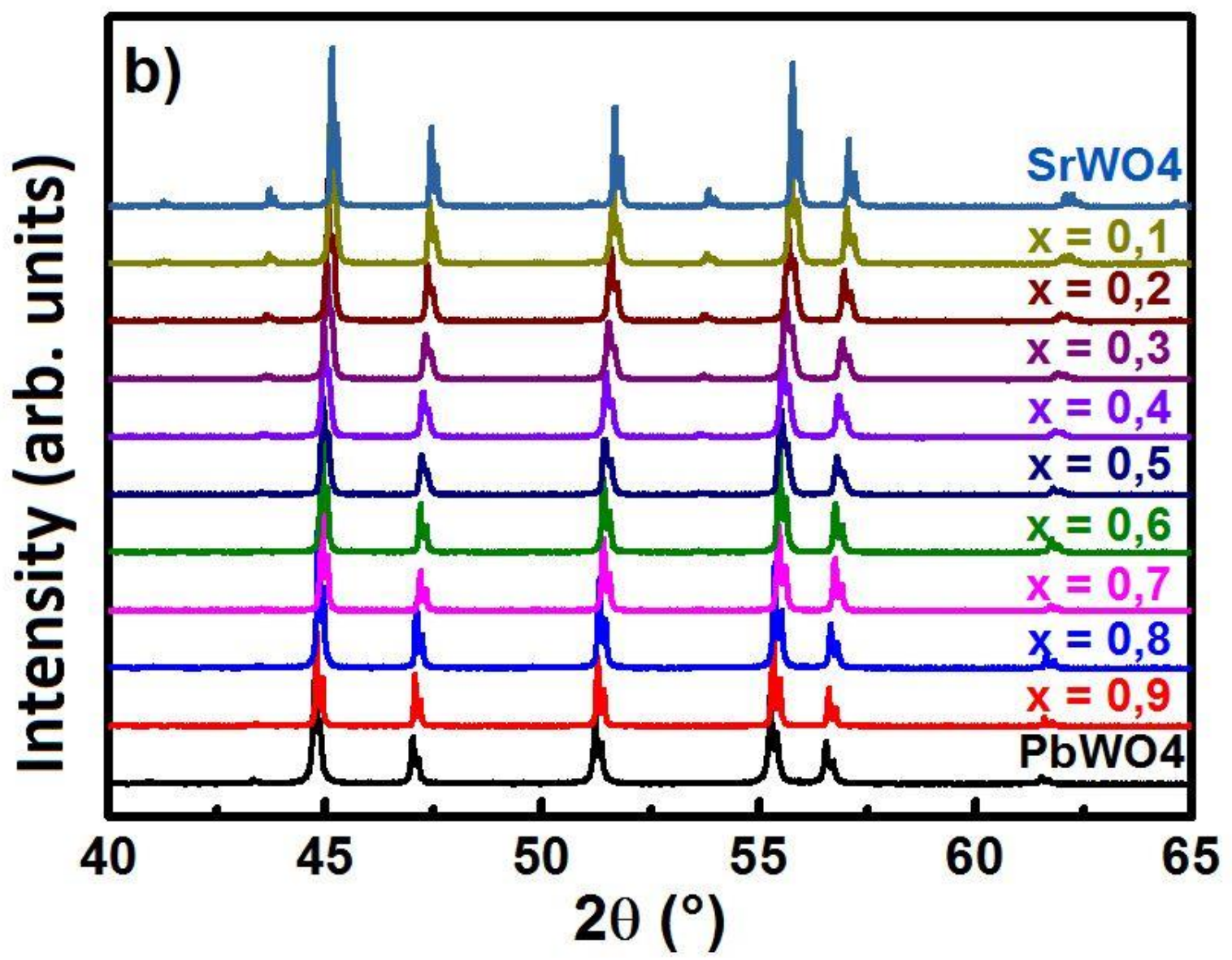

**Fig. 3a: Visualization of a limited part of X-ray diffraction patterns for the series $Sr_{1-x}Pb_xWO_4$[14] with x ranging between 0 and 1 with increments Δx = 0.1.**

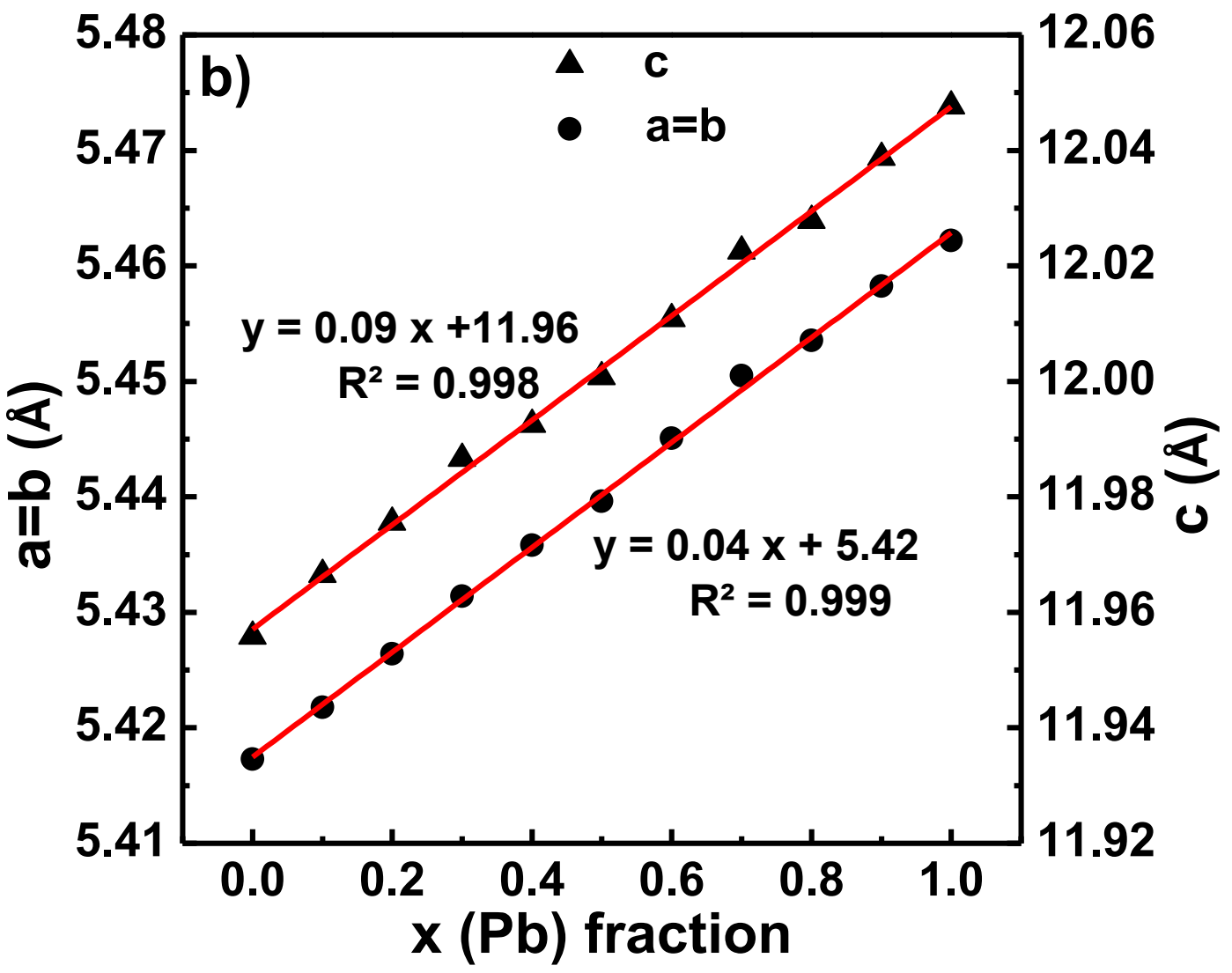

**Fig. 3b: Experimental crystal cell parameters of $Sr_{1-x}Pb_xWO_4$: variation as a function of x of a and c. Linear correlations have been drawn [14].**

### 3.1.2. Lattice distortions or size effects.

In the above Figure 3a, we can observe systematic broadening of each Bragg peak, with maximum values of FWHM's for $x_{max} = 0.35 \pm 0.05$. Considering the FWHM's representing the non-perturbed Bragg peaks of a well-crystallized standard sample, it is possible to determine the contribution Δ(FWHM) to the total FWHM of perturbed sample, susceptible to be interpreted in terms of lattice distortions:

$$(\varepsilon(x) = \langle \Delta a/a, (\Delta c/c)\rangle = \Delta\theta/\tan(\theta)$$

or size effects:

$$L(x) = 0.9\ \lambda\ /\ (\Delta 2\theta.\ \cos\theta).$$

These lattice distortions and local size effects might be separated using Williamson-Hall approach [18]: this separation between L and ε requires sufficiently large broadening of Bragg peaks, which is not the case for our materials. In fact, the SEM images of each sample (see ref. [14]) clearly showed that only large grains with crystallite sizes larger than 1 micron were present: this allowed us to exclude the formation of nanosized crystallites. For this reason, we have considered that these Δ(FWHM) variations could be described as pure distortions ε(x) with a good approximation. The lattice distortion function ε(x) reflect the presence of local zones with different cell parameters.

As a first approach, we have considered that the extreme phases have no distortion (or no size effect), and that the observed additional distortions ε(x) are mainly due to substitution and synthesis conditions. Having regard to the very small broadenings of Bragg peaks as x varies, it was not possible to clearly evidence any size effects in our experiments.

The function $Y_\varepsilon(x)$ to be fitted to the experimental data ε (x) associated to each composition x takes the general form (5):

$$\mathbf{Y_\varepsilon (x) = \varepsilon_{AA}.(1-x).G_{AA} (x) + \varepsilon_{BB}.x.G_{BB} (x) + \varepsilon_{AB} .x.(1-x).G_{AB}(x)} \quad \mathbf{(6)}$$

**Figure 4** reports the observed and calculated distortions resulting from the analysis of FWHM's of Bragg peaks with the fitting parameters $\varepsilon_{AA} = \varepsilon_{BB} = 0$, $\varepsilon_{AB} = 0.005$, $x_{(AB)} = 0.35$, $k_{(AB)} = 20$. These values could mean that distortions induced by substitution would be distributed with a typical characteristic $\delta x_{(AB)}$ of the gaussian function: $k_{(AB)} = (1/\delta x_{(AB)})^2$ with $\delta x_{(AB)}= 0.22$. The experimental value x=0.35 for which maximum of distortions effects occurs, coupled with this $\delta x_{(AB)} = 0.22$ value, would mean that the observed distortions should be mainly due to prominent contributions of AB mix zones.

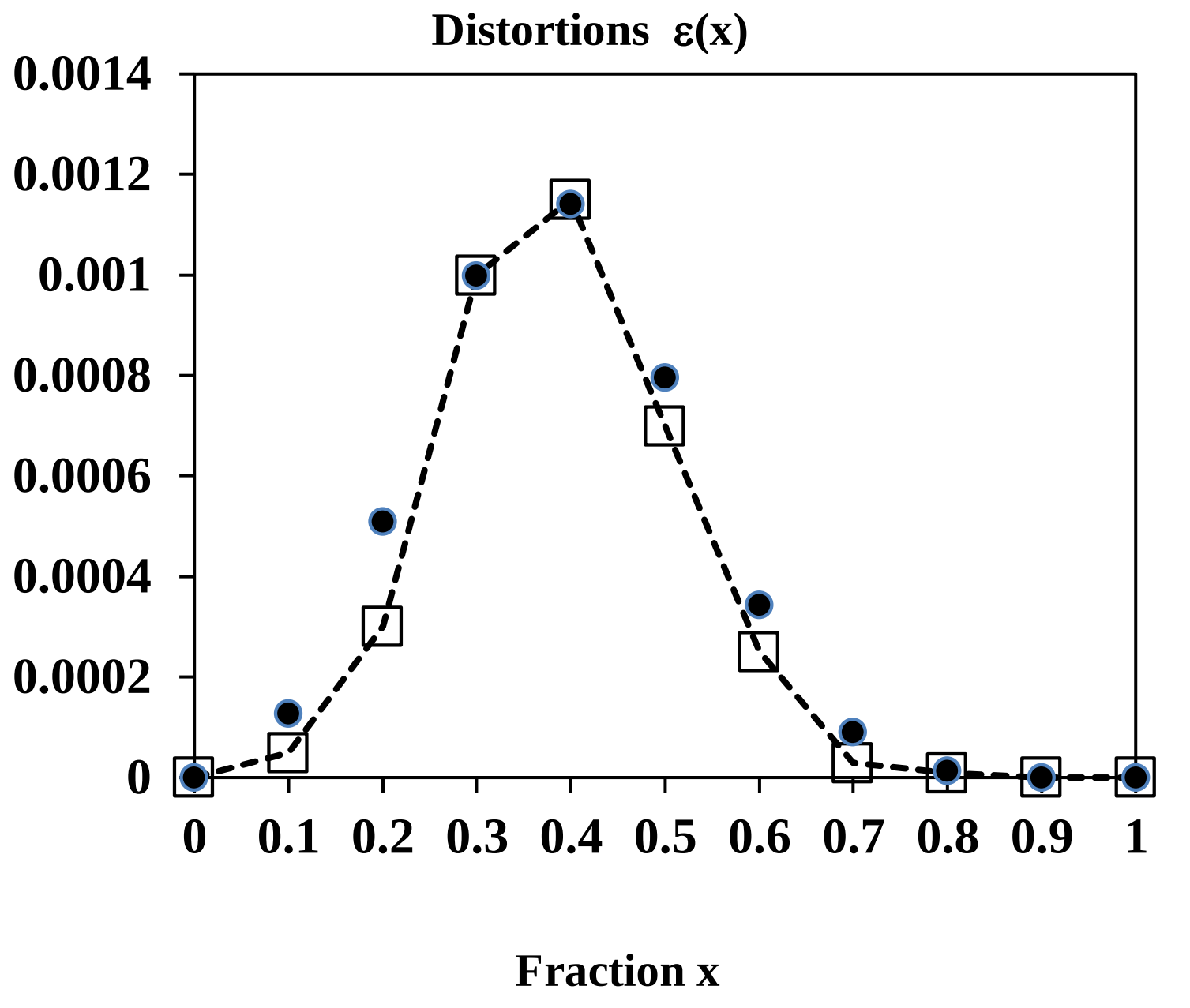

**Fig. 4: Observed (square) and calculated (round) distortion functions $\varepsilon_{obs}(x)$ and $Y_\varepsilon(x) = \varepsilon_{calc}(x)$. Parameters: $\varepsilon_{AA} = \varepsilon_{BB} = 0$ ; $\varepsilon_{AB} = 0.005$ ; $x_{AA} = 0$; $x_{BB} = 1$ ; $x_{AB} = 0.35$ ; $k_{AB} = 20$.**

**3.1. 3. Debye-Waller factors ($B_{DW}$).**

In the case of $B_{DW}$ factors, the X-ray diffraction analysis can deliver a determination of average displacements due to static disorder and thermal amplitudes of vibration: each AA, BB, AB zone can be perturbed with distribution of distortions. However, the determination of the static contribution $B_{DW(st)}$ of these $B_{DW}$ factors is not easy and can be obtained with high uncertainties. We have simulated a specific variation of this $B_{DW(st)}$ factor (ref. [14]) in clear correspondence with the experimental data shown in Fig. 5:

$$\mathbf{Y_{DW}(x) = B_{DWAA}.(1-x).exp(-k_A(x)^2)) + B_{DWBB}.x.exp(-k_B(1-x)^2)) + B_{DWAB}.x.(1-x).exp(-k_{AB}(x-x_{AB})^2))} \quad \mathbf{(7)}$$

The simulation was obtained with the following fitting parameters: $B_{DW(AA, BB, AB)}$ = 0.37/0.38/1.7; $x_{(A,B,AB)}$ = 0/1/0.5 ; $k_{(A,B,AB)}$ = 0/0/50.

Despite the uncertainties on the experimental values of $B_{DW}$ factors, a reasonable fit of calculated function to the observed data was obtained with a maximum perturbation $\mathbf{Y_{DW}(x)}$ for x=0.5. These fitting parameters could express a maximum disorder in AB zones, induced by disordered distributions of $Sr^{2+}$ and $Pb^{2+}$ cations with different sizes.

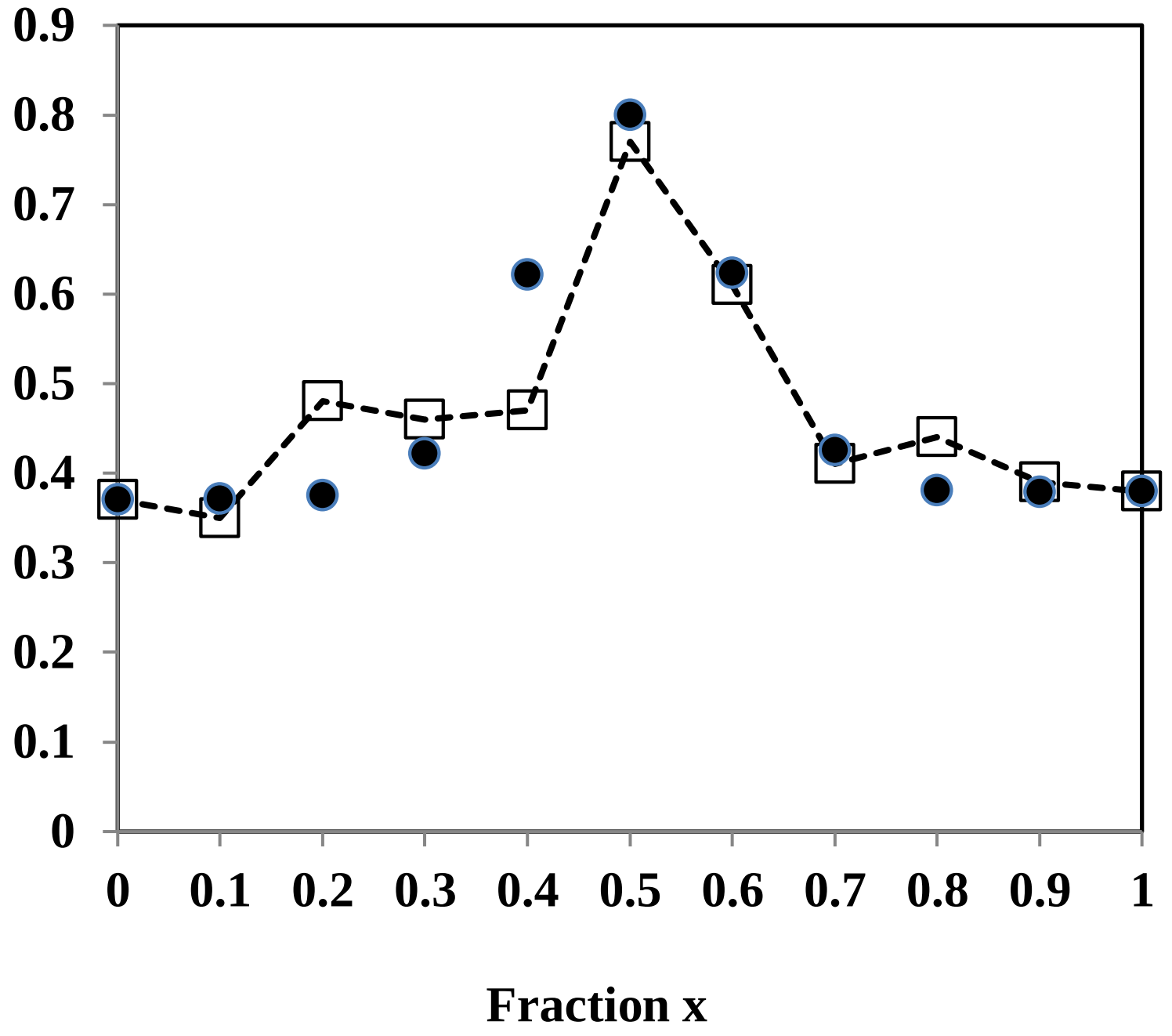

**Fig. 5: Experimental (square) $B_{DW}$ and calculated (round) $Y_{DW}$ Debye-Waller factors, as a function of composition x. Parameters: $B_{DW(AA,BB,AB)}$ = 0.37/0.38/1.7; $x_{(A,B,AB)}$ = 0/1/0.5 ; $k_{(A,B,AB)}$ = 0/0/50 .**

### 3.2. Modifications in Raman spectra.

Raman spectra of the tungstate series $Sr_{1-x}Pb_xWO_4$ were recorded and interpreted. **Figure 6** reports the experimental Raman bands for variable compositions. The high wavenumbers are associated with Ag, Bg, Eg modes characteristic of vibrations in $WO_4$ groups. The Ag mode close to 920 $cm^{-1}$ is characterized by a decreasing wavenumber as x increases, meaning that the $WO_4$ groups are sensitive to Pb-O-W interactions, with a softening of chemical bonds W-O.

The 920 $cm^{-1}$ Ag band is also characterized by a broadening in the composition range $0.3<x<0.8$ indicating the presence of distortions or size effects in the lattice. The main feature presently discussed through the model resides in the strong modification of the couple of Eg-Bg modes (800 – 850 $cm^{-1}$). In the composition range $0.2<x<0.8$, corresponding to a maximum distortion observed in X-ray diffraction patterns for x = 0.3-0.4, the Eg mode appears as being split into two modes progressively, with progressive evolution of intensities of these small bands as x increases. This can be clearly ascribed to the existence of two different environments $SrO_8$ / $PbO_8$, with an increasing proportion of $PbO_8$ groups perturbing the $WO_4$ tetrahedral groups.

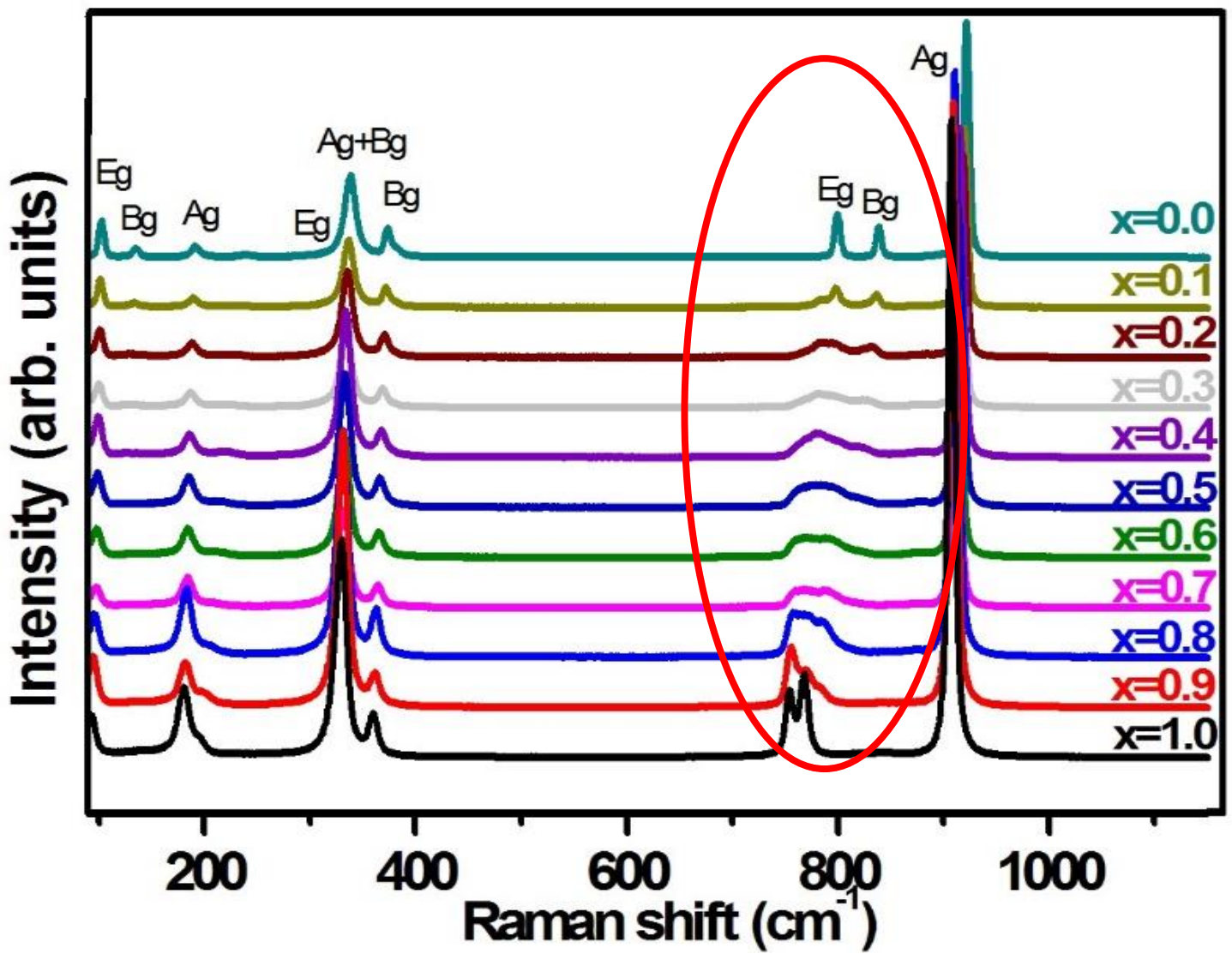

**Fig. 6: Raman spectra of SrPbWO series. Modification of the Eg and Bg modes as x varies.**

**Figure 7** shows the variations with composition x of the Ag, Bg and Eg modes. In **Figure 8**, we have simulated the variation of the separation of Eg mode and compared the calculated values to the experimental ones. As the wavenumbers of Ag and Bg modes decrease linearly with x, no simulation was necessary.

The separation **Δν** of the Eg split mode can be expressed through the function $\mathbf{Y_{\Delta\nu}(x)}$ as follows:

$$\mathbf{Y_{\Delta\nu}(x) = \Delta\nu_{AA}.(1-x).G_{AA}(x) + \Delta\nu_{BB}.x.G_{BB}(x) + \Delta\nu_{AB}.x.(1-x)\ G_{AB}(x)} \quad \mathbf{(8)}$$

A simulation was carried out with the following parameters: $\Delta\nu_{(AA,BB,AB)}$ = 0/0/120 (in $cm^{-1}$) ; $x_{(AA,BB,AB)}$ = 0/1/0.5 ; $k_{(AA,BB,AB)}$ = 0/0/5. The maximum of separation is observed for a composition of about $x_{AB}$=0.5, close to the $x_{AB}$ value calculated for the function $Y_{DW}(x)$, but slightly shifted from the maximum position observed for $Y_{\varepsilon}(x)$ at x = 0.35.

It should be remarked that the agreement between calculated values and experimental data cannot be optimal because of the uncertainties of wavenumber determinations.

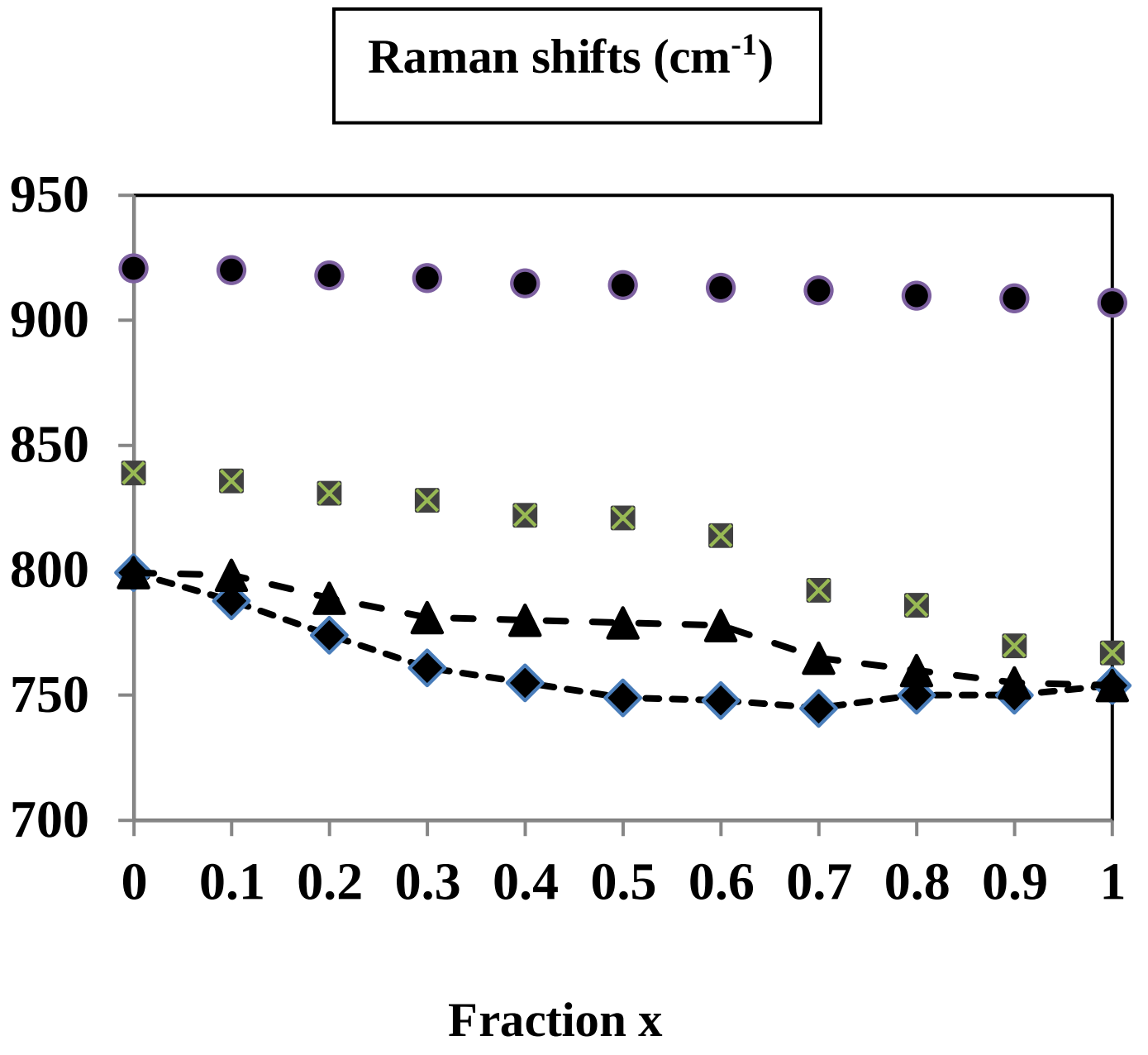

**Fig. 7: Experimental variations of Raman shifts (in $cm^{-1}$), as a function of x, for modes Ag(920), Bg(840) and the Eg modified modes at 800 $cm^{-1}$. Maximum of separation of Eg-like doublet observed for x= 0.55.**

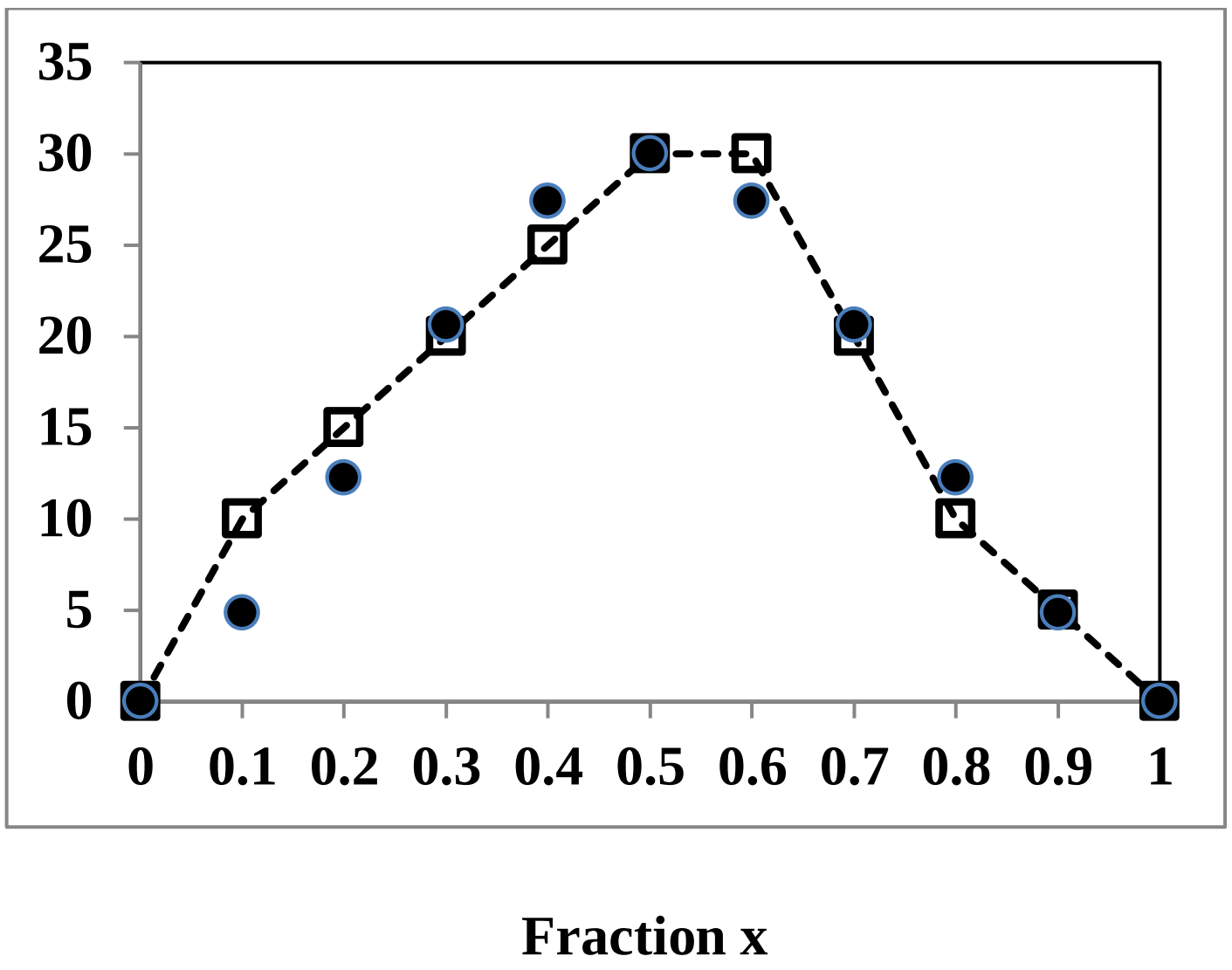

**Fig. 8: The Eg split mode: variations Δν(x). Experimental (square) and simulated (round) separation Δν(x). Maximum for x=0.55. $\Delta\nu_{(AA,BB,AB)}$ = 0/0/120; $x_{(AA,BB,AB)}$ = 0/1/0.5 ; $k_{(AA,BB,AB)}$ = 0/0/5.**

### 3.3. Modifications of Photoluminescence (PL) emissions.

The photoluminescence properties of each compound $Sr_{1-x}Pb_xWO_4$ depend on energy gaps of the extreme phases $SrWO_4$ (Eg close to 5 eV) and $PbWO_4$ (Eg close to 4 eV). The PL properties of these Sr and Pb tungstates were extensively studied by the past. Their electronic band structures were determined [3]. In the specific case of $PbWO_4$, it was shown that the Pb6s orbitals played an important role in the valence band corresponding to the O2p orbitals. The calculations by authors [3] showed that the Pb6s-O2pσ* molecular orbitals would contribute to state densities at the top of the valence band (see **Figure 9** below), which is essentially based on hybrid O2p orbitals. Consequently, a significant modification of the emissive properties, but also of absorption under photonic excitation, can be expected when the composition in lead increases.

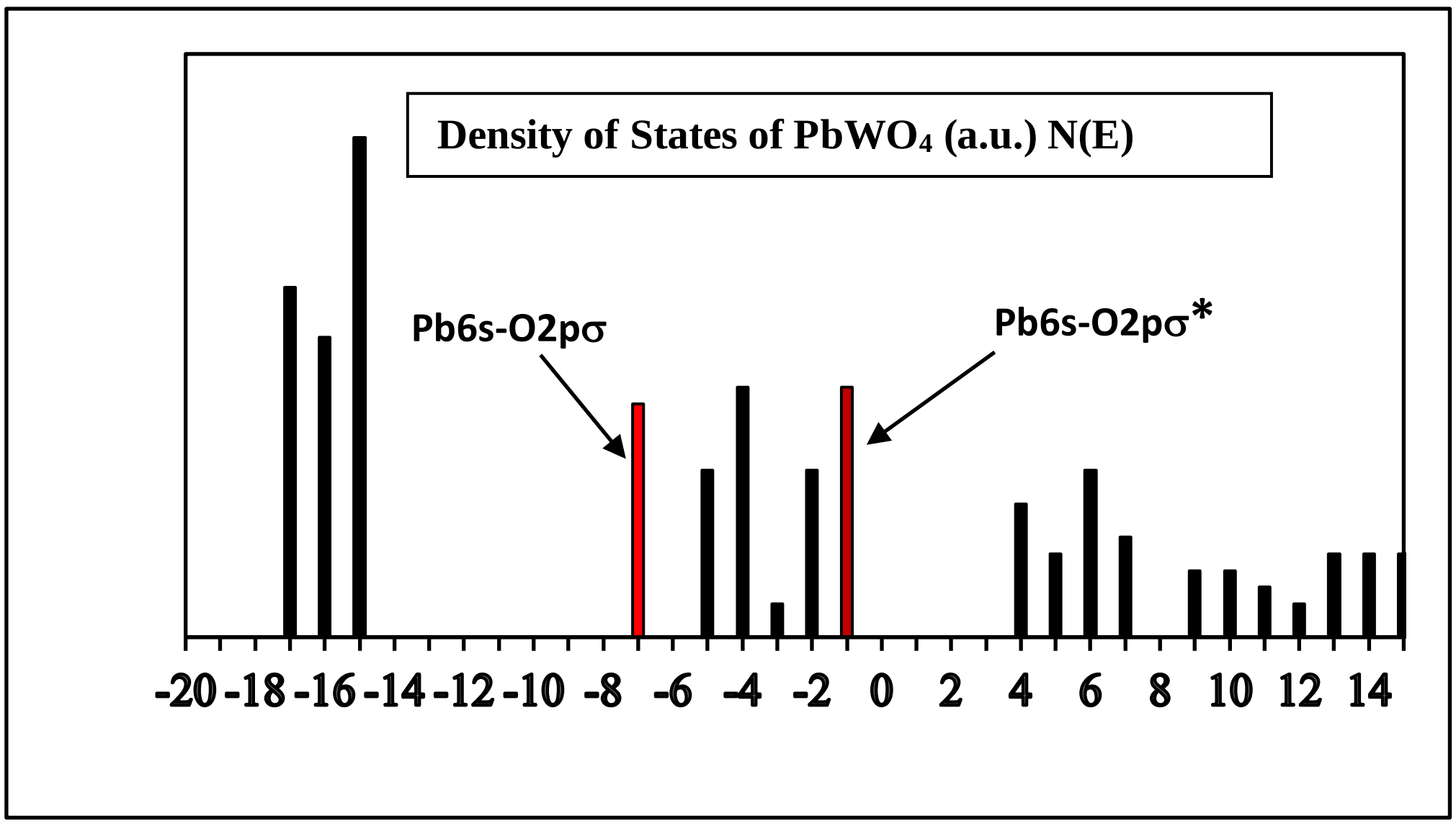

Energy (eV)

**Fig. 9: Simplified representation of electronic band structure of $PbWO_4$ according to reference [3]: density of states N(E) with E in eV. Valence band (VB = -6 to 0 eV); conduction band CB >4 eV; band gap Eg = 4 eV. Presence of Pb orbitals at the top of VB.**

In the case of the $Sr_{1-x}Pb_xWO_4$ and $Sr_{1-x}Pb_xMoO_4$ series, experimental PL spectra were obtained in two types of experiments**:**

(i) under X-Ray excitation using the global emission from copper source (E<45 keV) [14,15];

(ii) under UV laser excitation (E = 3.4 eV or 364.5 nm wavelength)

Figures 10a and 10b report the various spectra for x ranging between 0 and 1 under X-rau excitation (10a) and under UV excitation (10b). Figure 10c reports three typical spectra obtained under X-ray excitation for compositions x=0, 0.3 and 1. Figure 10d reports typical spectra obtained under UV excitation for compositions x= 0, 0.7 and 1. Figures 10e and 10f give the variations of spectral intensities as a function of composition x under X-Ray and UV excitations respectively.

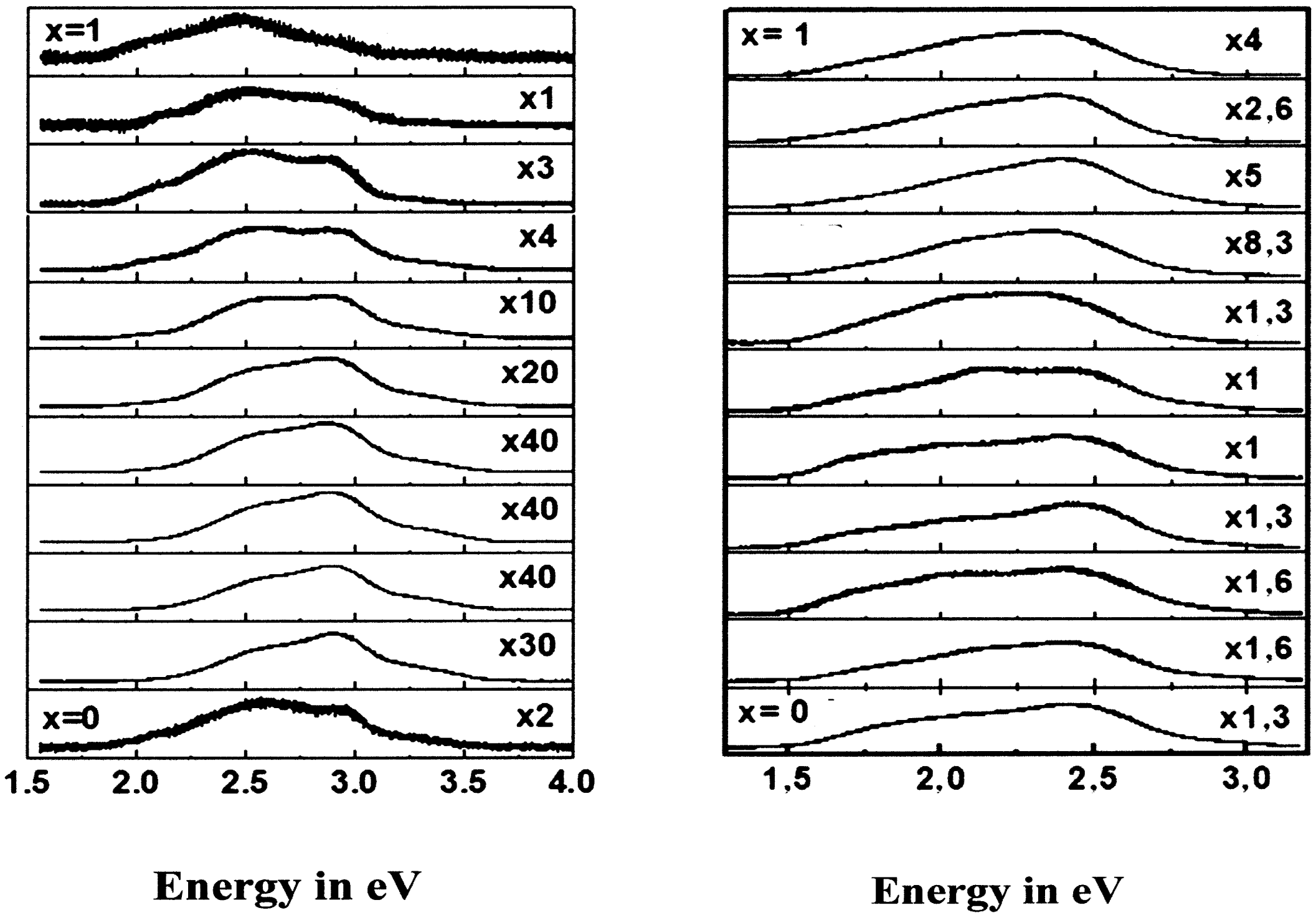

**Fig. 10a,b: Photoluminescence intensities of $Sr_{1-x}Pb_xWO_4$. $I_{PL}$/XR(x) under XRD excitation -(a) = left Figure-, and IPL/UV under UV excitation -(b) = right Figure-. Three or four gaussian components allow reproducing the full PL spectra as composition x varies ($0 \leq x \leq 1$, with steps of $\Delta x = 0.1$).**

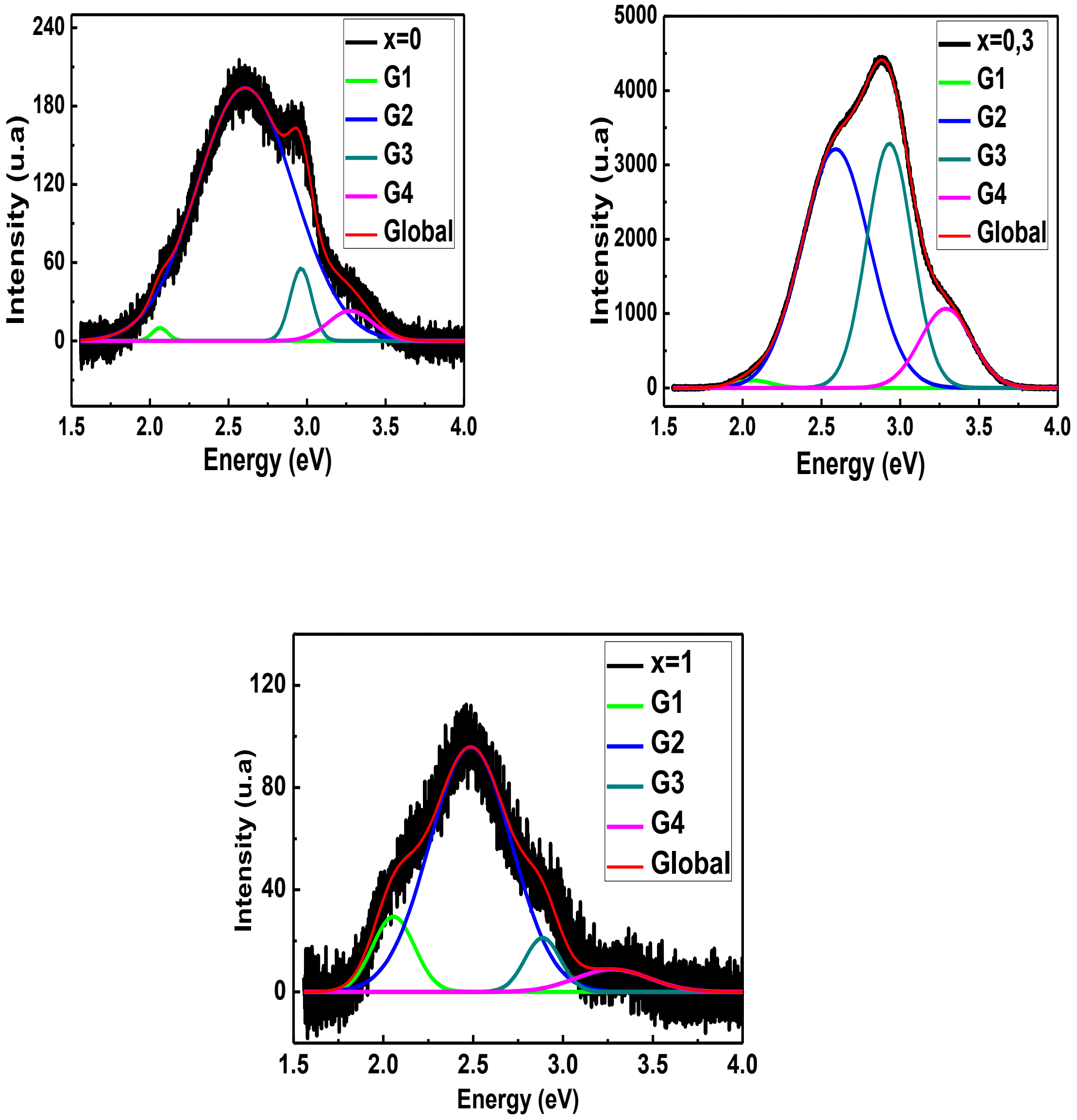

**Fig. 10c: PL spectra (in electronic shots) of $Sr_{1-x}Pb_xWO_4$ phases, under X-ray excitation, for compositions x=0, 0.3, 1. The maximum intensity was reached for composition x=0.3. Each spectrum can be decomposed into 3 or 4 gaussian components (see [15]). Two main components at 2.55 and 3 eV are accompanied by small components generated by defects and distortions.**

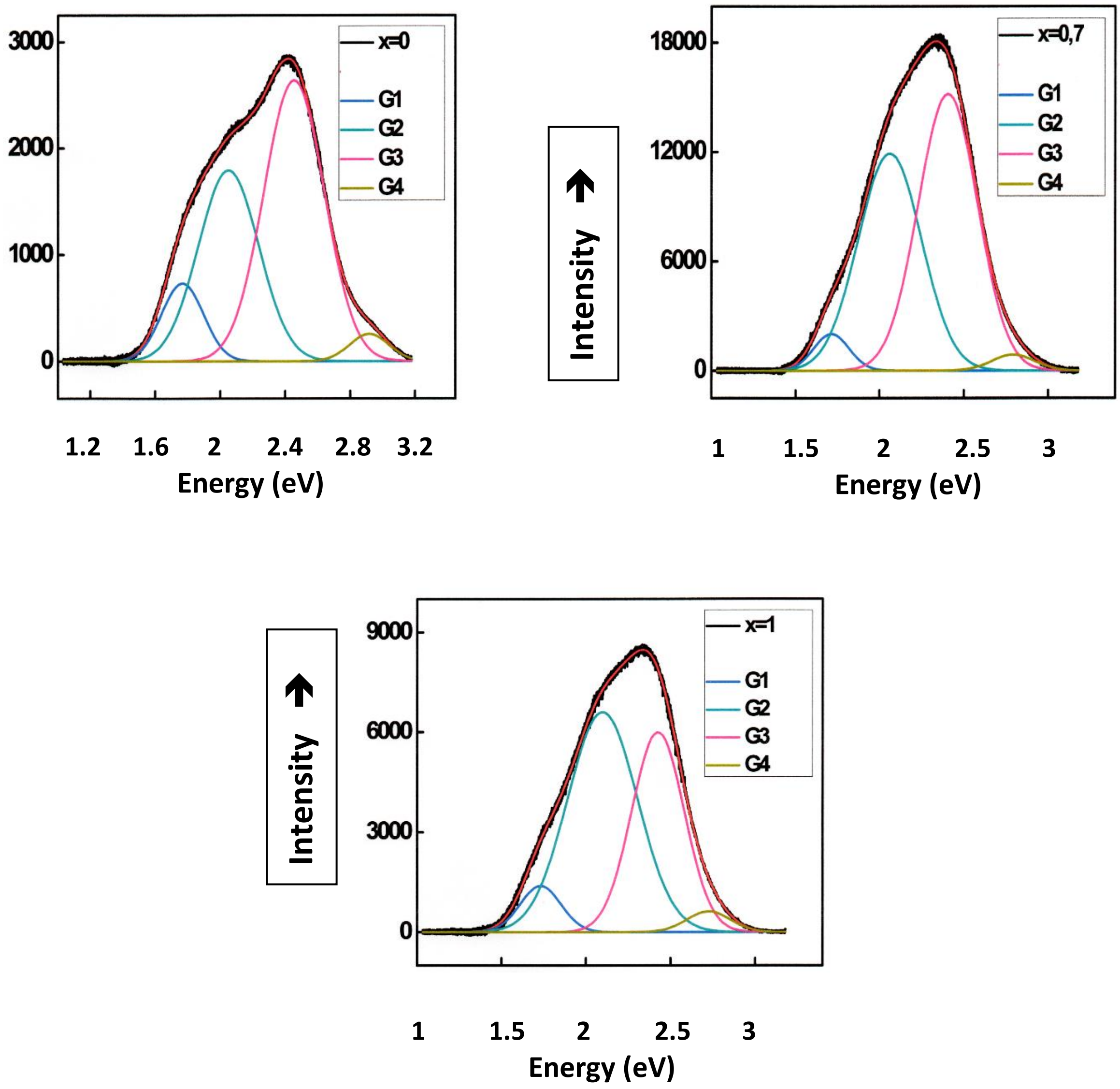

**Fig. 10d: PL spectra (in electronic shots) of $Sr_{1-x}Pb_xWO_4$ phases, under UV excitation, for compositions x=0, 0.7, 1. The maximum intensity was reached for composition x=0.7. Each spectrum can be decomposed into 3 or 4 gaussian components (see [17]). Two main components with energies close to 2 and 2.5 eV are accompanied by smaller components generated by surface defects.**

Under high energy polychromatic X-ray excitation, the maximum intensity was observed close to $x_{max} = 0.3 \pm 0.1$ in both series. All spectra were decomposed into four emissive components.

Two components were strongly correlated with the substitution rate x. Under monochromatic UV excitation (with wavelength of 364.5 nm), for both series of materials, the emitted spectra were characterized by a significant increase of the emitted intensity for a composition close to $x_{max}$ = 0.7 ± 0.1. To date, the exact origin of these intensity maximum, under X-ray or UV excitation, has remained without complete explanation. In Figures 10e and 10f, the variations of PL intensities as a function of composition x are reported showing the different maxima observed under X-ray and UV excitations.

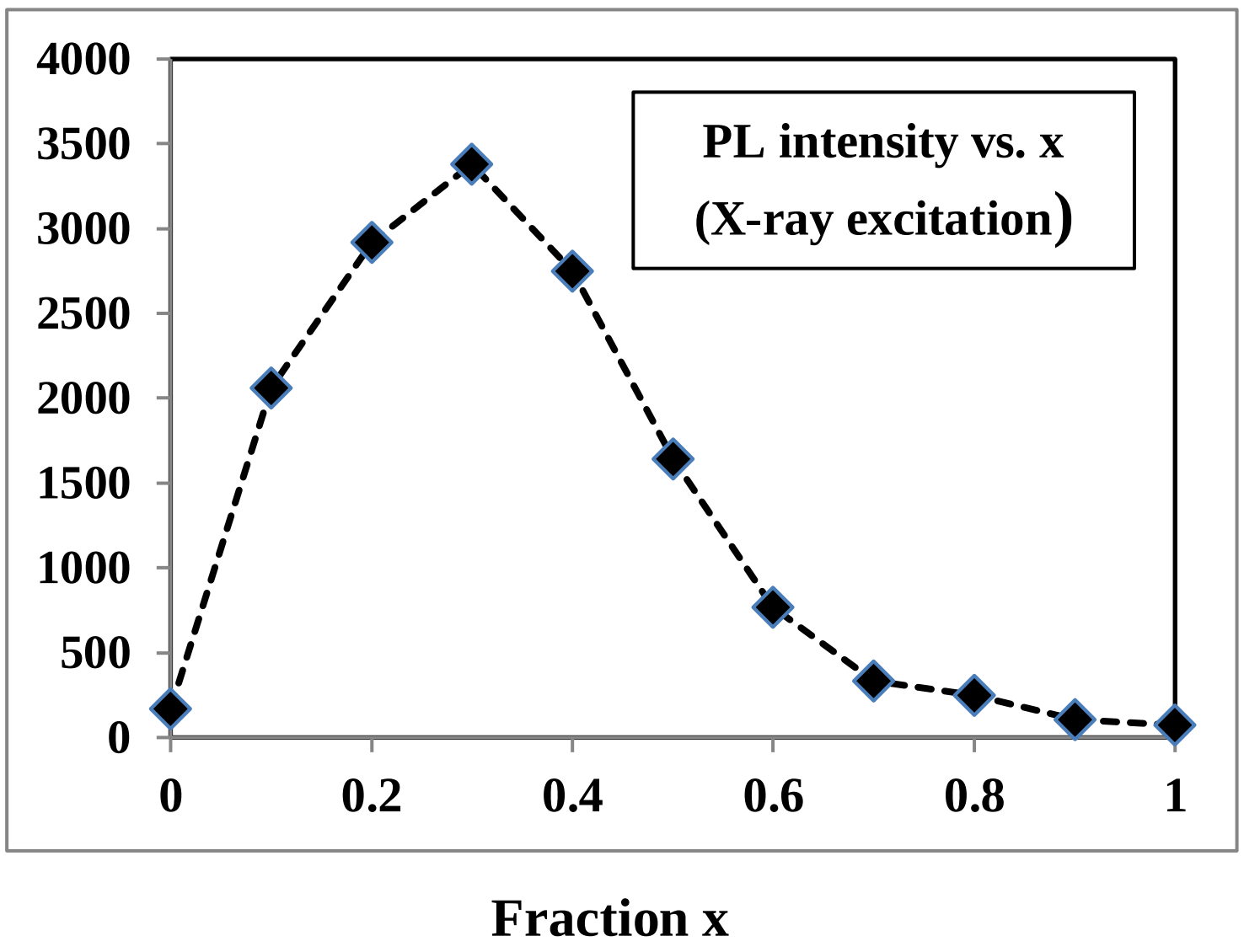

**Fig. 10e: Variation of total PL intensities of $Sr_{1-x}Pb_xWO_4$ polycrystalline samples subjected to X-ray excitation (Cu polychromatic source), as a function of composition x. Maximum of PL emission at x = 0.30 .**

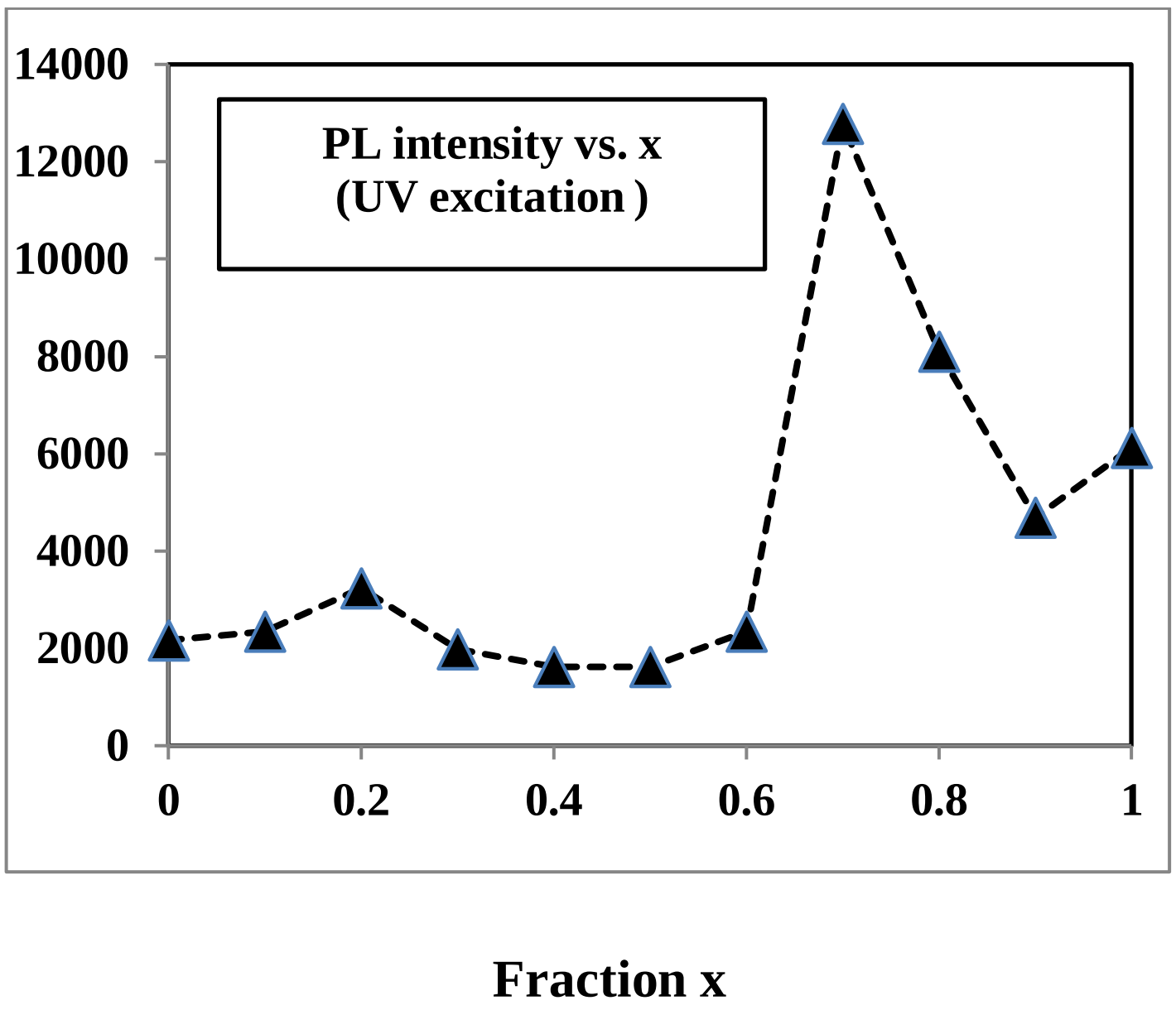

**Fig. 10f: Variation of total PL intensities of $Sr_{1-x}Pb_xWO_4$ polycrystalline samples, subjected toUV excitation (364.5 nm), as a function of composition x. Maximum of PL emission at x = 0.70. (from Hallaoui's thesis).**

In the case of X-ray excitation, a simulation of the variation with composition x of intensities $Y_{PL}(x)$ was carried out using the following form:

$$\mathbf{Y_{PL}(x) = Y_{PLAA}.(1-x).\exp(-k_{AA}(x)^2)) + Y_{PLBB}.x.\exp(-k_{BB}(1-x)^2)) + Y_{PLAB}.x.(1-x).\exp(-k_{AB}(x-x_{AB})^2))} \quad \mathbf{(9)}$$

where $Y_{PL(AA,BB,AB)}$, $x_{(AB)}$ and $k_{(AA,BB,AB)}$ are the appropriate fitting parameters for PL emission.

Figure 11 compares the experimental $I_{PL}(x)$ and simulated $Y_{PL}(x)$ values of PL intensities under X-ray excitation as a function of composition x. To fit calculated to experimental values of PL emissions, we have used the following parameters: $I_{PL(AA,BB,AB)}$ = (170/80/17000) ; $x_{(AA,BB,AB)}$ = (0 / 1/ 0.2) ; $k_{(AA/BB/AB)}$ = (4 /8/ 10).
These obtained values of parameters can be justified by the existence of:

- small number of active centers in AA and BB zones,
- significant number of centers in AB zones directly due to substitution. In addition, the values of the k parameters for AA and BB zones could mean that centers would be regularly distributed in the full composition range, while the k value for AB zones would mean that the distribution of centers would be narrower with δx= 0.32, with a maximum at x=0.30.

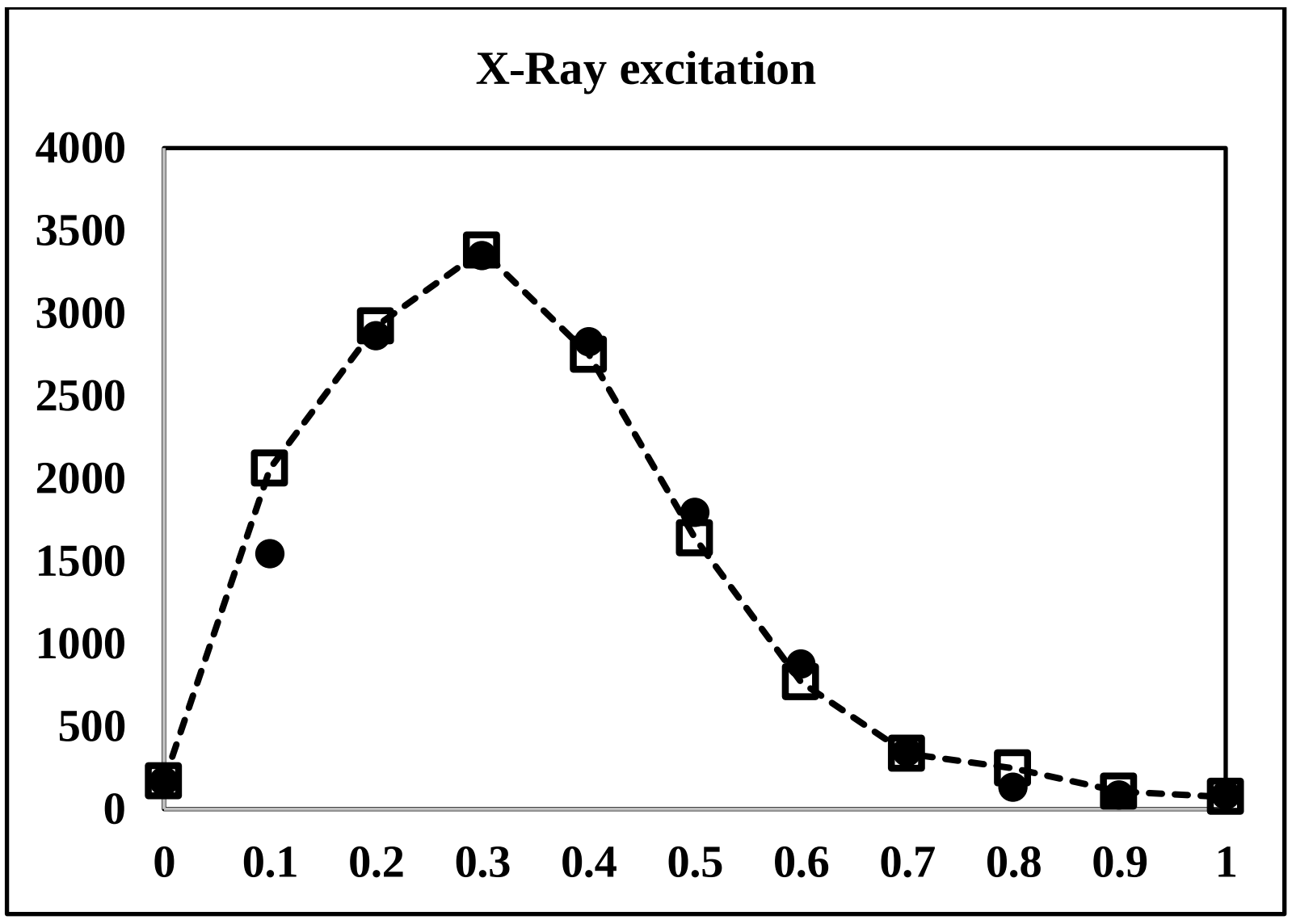

**Fig. 11: Experimental (square) $I_{PL}$ and calculated (round) $Y_{PL}$ intensities of PL bands in $Sr_{1-x}Pb_xWO_4$ under X-ray excitation. $Y_{PL(AA,BB,AB)}$ = (170/80/17000); $x_{(AA,BB,AB)}$ = (0/1/0.2); $k_{(AA/BB/AB)}$ = (4/8/10).**

Similar simulations of these characteristics were performed for this series using the $I_{PL}$ intensities obtained under UV excitation: in this case of low energy excitation, the maximum of PL intensity has been moved to a value x=0.7, fully different from the one obtained under X-ray excitation. **Figure 12** compares the $I_{PL}(x)$ to $Y_{PL}(x)$ values for UV excitations.

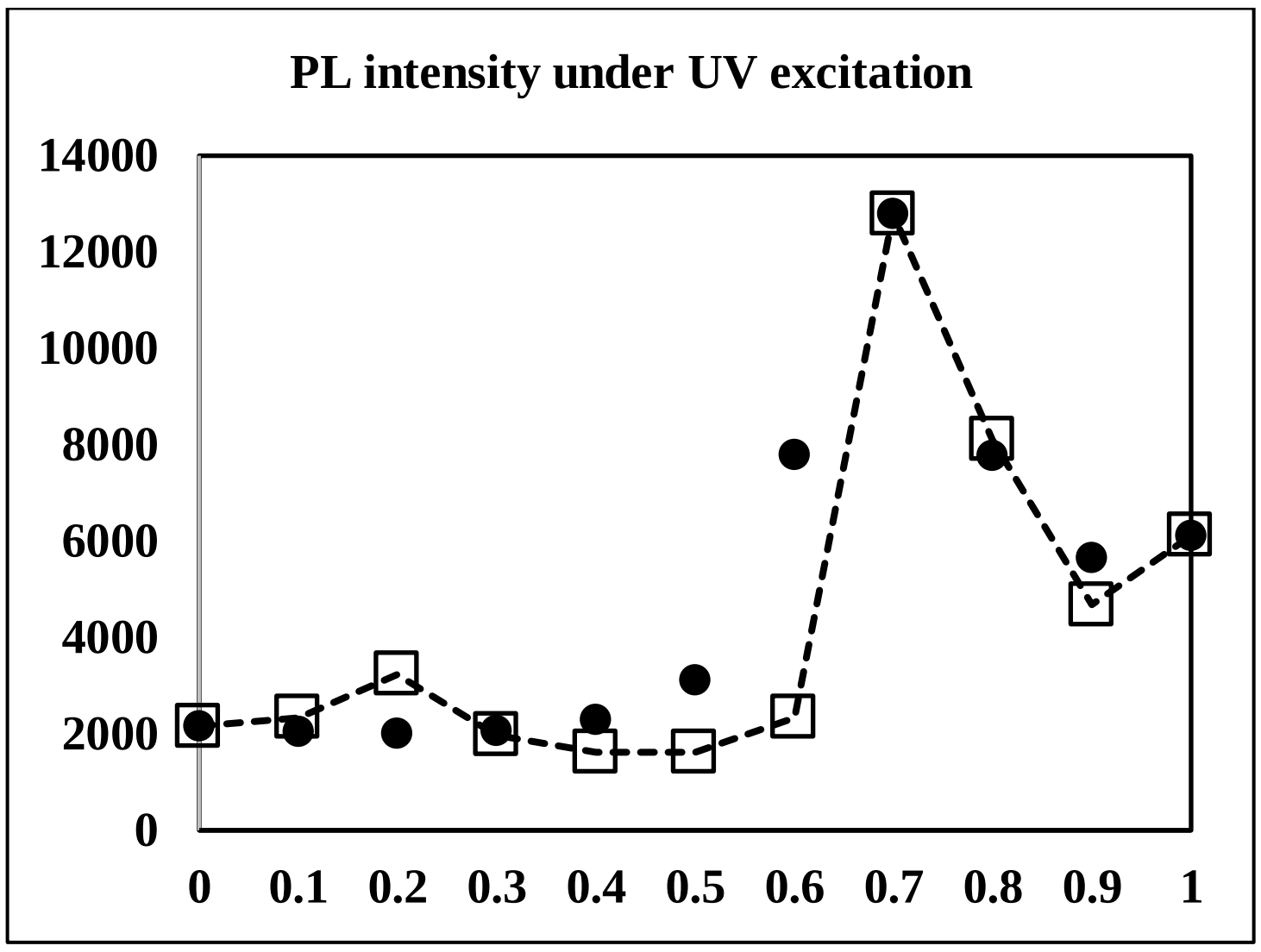

**Fraction x**

**Fig. 12: Experimental (white square) $I_{PL}$ and calculated (black round) $Y_{PL}$ intensities of PL bands in $Sr_{1-x}Pb_xWO_4$, under UV excitation. $Y_{PL(AA,BB,AB)}$ = (2160/6120/42000) ; $x_{(AA,BB,AB)}$ = (0/1/0.7) ; $k_{(AA/BB/AB)}$ = (2/1/80).**

These two different compositions x=0.3 (XR excitation, high energy) and x=0.7 (UV excitation, low energy) were ascribed to complex effects. The high-energy polychromatic X-ray irradiation can penetrate deeply the polycrystalline material and excite the bonding and antibonding "Pb6s" levels located at the top and just below the valence band constituted by oxygen O2p orbitals. The low energy UV laser irradiation can interact only with the surface of material and could excite only the top of the valence band. This could be at the origin of the two different intensity maxima observed for the compositions x= 0.3 for XR excitation, and x= 0.7 for UV excitation.

## 4. Discussion.

In the **Table** below, we have reported the various parameters corresponding to each property. The $x_{max}$ value is the composition corresponding to the maximum of the experimental function $I_{PL}(x)$. The calculated $x_{AB}$ values corresponding to the $Y_{PL}(x)$ function can slightly differ from the experimental $x_{max}$ ones: this is a result of fitting Y to P data.

**Table: Simulation parameters for the $Sr_{1-x}Pb_xWO_4$ series.**

| Y(x) | $Y_{AA}$ | $Y_{BB}$ | $Y_{AB}$ | $x_{AA}$ | $x_{BB}$ | $k_{AA}$ | $k_{BB}$ | $k_{AB}$ | $x_{AB}$ | Experimental $x_{max}$ |
|---|---|---|---|---|---|---|---|---|---|---|
| **a(x)** | 5.414 | 5.461 | 0 | | | | | | | |
| **c(x)** | 11.93 | 12.055 | 0 | | | | | | | |
| **ε(x)** | 0 | 0 | 0.005 | 0 | 1 | 0 | 0 | 30 | 0.35 | **0.35** |
| **$B_{DW}$(x)** | 0.37 | 0.38 | 1.7 | 0 | 1 | 0 | 0 | 50 | 0.50 | **0.50** |
| **Δν(x)** | 0 | 0 | 120 | 0 | 1 | 0 | 0 | 5 | 0.50 | **0.55** |
| **$I_{PL}$(x)/XR** | 170 | 80 | 17000 | 0 | 1 | 4 | 8 | 10 | 0.20 | **0.30** |
| **$I_{PL}$(x)/UV** | 2160 | 6120 | 42000 | 0 | 1 | 1 | 2 | 80 | 0.7 | **0.7** |
| Definitions: a, b, c (x) = cell parameters in Å; ε(x) = average lattice distortion parameter; $B_{DW}$(x) = Debye Waller factor (in Å$^2$); Δν (x) = Raman separation of initial Eg mode in cm$^{-1}$; $I_{PL}$ (x)/XR = PL intensity under X-ray excitation in arb. unit; $I_{PL}$ (x)/UV = PL intensity under UV excitation in arb. unit. | | | | | | | | | | |

The $Y_{(AA,BB)}$ values are characteristic of unmodified compounds (x=0 and 1). The $G_{(AA,BB)}$ functions have been introduced to take into account possible modifications of pure local zones due to interactions with neighboring zones. The $Y_{AB}$, $x_{AB}$ (or $x_{max}$) and $k_{AB}$ parameters play the major role in the modifications involved by substitution: the resulting $G_{AB}$ function represent the extension of modification close to the $x_{max}$ value.

**Lattice distortions.** The lattice perturbation function linked to diffraction profiles ε(x) (average values between Δa/a and Δc/c variations) is characterized by a unique $Y_{AB}= \varepsilon_{AB}$ value, resulting from the choice of absence of distortion for extreme phases. The observed maximum at $x_{max}$ = 0.35 (±0.10) with extension parameter k of 30 (0.18 in composition) could mean that the maximum of lattice distortions should be reached at this composition.

**Local disorder**. In the case of $Y_{DW}$(x) (DW factors) a maximum of static displacements is observed at $x_{max}$ = 0.5 (±0.15): having regard to the uncertainties in DW factor determinations, we can conclude to an acceptable agreement with the $x_{max}$ found for ε(x).

**Perturbation of Raman Eg modes.** The $Y_{\Delta\nu}$(x) function is also characterized by a unique $Y_{AB}$ parameter: the observed maximum is close to $x_{max}$ = 0.55. It clearly characterizes the presence of AA, BB and AB zones, with Eg modes characterizing each of these zones.

**Maximum of Photoluminescence under X-ray excitation.** In the case of X-ray excitation, the PL emissions would result from a complex distribution of defects and PL active centers. The maximum of $Y_{PL}$(x) observed under X-ray excitation at x = 0.30 corresponds to the observed maximum of lattice distortions ε(x) or minimum sizes (L(x) of local zones AB, AA and BB). The calculated $x_{AB}$ = 0.20 value allows a better fit of the $Y_{PL}(x)$ function and is strongly correlated to the contributions $Y_{BB}$ and $Y_{AB}$ (see Table) stronger than the contribution $Y_{AA}$. In other words, the increasing fraction of Pb should be associated with increasing population of PL centers in the solid solution and would be at the origin of the maximum of PL emission. The k values suggest that small perturbations of intrinsic emission centers occur in the AA and BB zones, while the distribution of emission centers should be centered around x= 0.3 and narrower.

**Maximum of Photoluminescence under UV excitation.** In the case of UV excitation, the PL emissions present a maximum shifted to $x_{max}$ = 0.7 (±0.1). The high $k_{AB}$ value means that the distribution of emission centers due to the formation of links Sr-W-O-Pb is narrower than in the case of X-ray excitation. This can be related to the weaker penetration of UV beam in the bulk and to emissions from perturbed (distorted) $WO_4$ tetrahedra located in the grain surfaces: these surfaces should present emission centers differing from the emission centers in the bulk.

## 5. Conclusions.

The as-proposed semi-empirical model was applied to different properties of $Sr_{1-x}Pb_xWO_4$ solid solution: it delivered a series of parameters characteristic of the perturbations of properties resulting from substitution and synthesis conditions. For a given thermal treatment (fixed temperature and heating time), **the local zones AA, BB and AB** can present variable extensions and deformations as a function of composition x, giving rise to variable individual properties. The PL emission spectra obtained under X-ray or UV excitations are characterized by closely related energies corresponding to charge transfers in $WO_4$ groups perturbed by point defects and substitution. The two different maxima of intensities observed for two different compositions $x_{max}$ = 0.3 and 0.7, under X-ray and UV excitations respectively, are data introduced in the model, to determine first approximate values of fitted $x_{AB}$ values. They result from a compromise between increase of emission centers and increase of absorption as x increases. The pertinent parameters proposed by the model to simulate various properties are the 3 parameters $Y_{(AA,BB,AB)}$, the 3 parameters $k_{(AA,BB,AB)}$ and the $x_{AB}$ value closely related to $x_{max}$.

These seven fitting parameters must be considered as only indicative of the state of disorder and deviations from ideal material $A_{(1-x)}B_{(x)}XO_4$, with specific properties, more or less closely related to linear combinations of properties of the extreme phases $AXO_4$ and $BXO_4$. Each parameter is characteristic of the history of the sample and of substitution. This approach applied to the strontium lead molybdate series gave similar results. Now, this approach is going to be applied to various series of solid solutions.

**Acknowledgements:** This study results from international cooperation between IM2NP, University of Toulon, University Ibn Zohr (Agadir) during years 2010 to 2018. It follows a first project (Nanogamma) involving notably the CEA of Cadarache and CESIGMA of La Garde. Financial supports were obtained from the Regional Council of Provence-Alpes-Côte d'Azur (no. 2012-16322), of the General Council of Var, of Toulon Provence Mediterranean, and from part of the CNRS-CNRST project (Chimie 02/14).

## References.